\documentclass[manuscript,acmsmall,screen]{acmart}

\AtBeginDocument{%
  \providecommand\BibTeX{{%
    \normalfont B\kern-0.5em{\scshape i\kern-0.25em b}\kern-0.8em\TeX}}}

\usepackage{booktabs} 

\usepackage[ruled]{algorithm2e} 
\usepackage{graphicx}
\usepackage{textcomp}
\usepackage{color,soul}
\usepackage{bm}
\usepackage{multirow}
\usepackage{subcaption}
\usepackage{balance}
\usepackage{graphicx}  

\usepackage{capt-of}
\usepackage{colortbl}
\usepackage{arydshln}
\usepackage{longtable,booktabs}
\usepackage{enumitem}

\definecolor{green}{rgb}{0.0, 0.65, 0.31}
\definecolor{bleudefrance}{rgb}{0.19, 0.55, 0.91}
\definecolor{ceruleanblue}{rgb}{0.16, 0.32, 0.75}
\definecolor{grey}{HTML}{969696}
\definecolor{violet}{HTML}{8856a7}
\definecolor{dgrey}{HTML}{01665e}
\definecolor{lgrey}{HTML}{5ab4ac}
\definecolor{dgreen}{HTML}{005a32}
\definecolor{purple}{HTML}{ae017e}
\definecolor{NewBlue}{HTML}{253494}
\definecolor{lblue}{HTML}{decbe4}

\definecolor{editCol}{HTML}{000000}
\definecolor{maskCol}{HTML}{c51b7d}
\definecolor{lrColor}{HTML}{8856a7}
\definecolor{trColor}{HTML}{d01c8b}
\definecolor{ctColor}{HTML}{4dac26}
\definecolor{brickred}{HTML}{f03b20}
\definecolor{improveCol}{HTML}{253494}
\definecolor{worsenCol}{HTML}{d7191c}
\definecolor{lgreen}{HTML}{e0f3db}
\definecolor{dpink}{HTML}{CD1076}
\definecolor{pink}{HTML}{FED2D2}
\definecolor{soothinggreen}{HTML}{4dac26}
\definecolor{darkred}{HTML}{8B0000}

\definecolor{dblue}{HTML}{104E8B}
\definecolor{violet}{HTML}{8A2BE2}
\definecolor{mscolor}{HTML}{01665e}
\definecolor{nmscolor}{HTML}{d8b365}
\definecolor{deepgrey}{HTML}{525252}
\definecolor{dslate}{HTML}{2F4F4F}
\definecolor{dolive}{HTML}{556B2F}
\definecolor{teal}{HTML}{388E8E}
\definecolor{mscolor}{HTML}{01665e}
\definecolor{nmscolor}{HTML}{d8b365}

\definecolor{srcolor}{HTML}{e34a33}
\definecolor{smcolor}{HTML}{253494}
\definecolor{srsmcolor}{HTML}{7fcdbb}
\definecolor{bothcolor}{HTML}{fe9929}
\definecolor{onecolor}{HTML}{018571}

\definecolor{beet}{HTML}{8E388E}

\newcommand{\tabitem}{\textbullet~~}

\newcommand*{\textlabel}[2]{%
  \edef\@currentlabel{#1}
  \phantomsection
  #1\label{#2}
}

\definecolor{yellowHL}{HTML}{ffffb3}
\definecolor{blueHL}{HTML}{bebada}
\definecolor{greenHL}{HTML}{8dd3c7}
\definecolor{orangeHL}{HTML}{fb8072}

\newcommand{\textqt}[1]{\small{#1}}

\newcommand{\SBSIE}[1]{\textsf{ESI}}
\newcommand{\SBJFP}[1]{\textsf{PJF}}
\newcommand{\SBWPS}[1]{\textsf{WPS}}
\newcommand{\SBNFT}[1]{\textsf{NFW}}

\colorlet{tableheadcolor}{gray!25} 
\colorlet{tablerowcolor}{gray!15} 
\colorlet{tablerowcolor2}{gray!12} 
\colorlet{tablerowcolor3}{gray!25} 
\colorlet{tablerowcolor4}{gray!50} 

\newcommand{\rowcollight}{\rowcolor{tablerowcolor2}} %

\newcommand{\edit}[1]{{\textcolor{editCol}{#1}}}

\renewcommand{\textrightarrow}{$\rightarrow$}

\colorlet{tableheadcolor}{gray!25} 
\colorlet{tablerowcolor}{gray!5} 

\definecolor{neutralCol}{HTML}{dd1c77}
\definecolor{neutralGreen}{HTML}{31a354}

\definecolor{bleudefrance}{rgb}{0.19, 0.55, 0.91}  
\definecolor{AfTrColor}{HTML}{0868ac}  
\definecolor{BfTrColor}{HTML}{a8ddb5}  

\definecolor{AfCtColor}{HTML}{b10026}  
\definecolor{BfCtColor}{HTML}{fd8d3c}

\graphicspath{ {figures/} }

\acmJournal{PACMHCI}
\acmYear{2023} \acmVolume{0} \acmNumber{CSCW2} \acmArticle{0} \acmMonth{8} 
\acmDOI{}

\setcopyright{none}

\usepackage{graphicx} 
\usepackage{subcaption}
\usepackage{makecell}
\usepackage{tabularx}
\usepackage{microtype}

\begin{document}

\title[Sensing Wellbeing in the Workplace, Why and For Whom?]{Sensing Wellbeing in the Workplace, Why and For Whom?\\ Envisioning Impacts with Organizational Stakeholders}

\author{Anna Kawakami}
\affiliation{%
  \institution{Carnegie Mellon University}
  \city{Pittsburgh}
  \state{PA}
  \country{USA}
}
\email{akawakam@andrew.cmu.edu}

\author{Shreya Chowdhary}
\affiliation{%
   \institution{University of Michigan}
  \city{Ann Arbor}
  \state{MI}
  \country{USA}
}
\email{schowdha@umich.edu}

\author{Shamsi T. Iqbal}
\affiliation{%
  \institution{Microsoft Research}
  \city{Redmond}
  \state{WA}
  \country{USA}
}
\email{shamsi@microsoft.com}

\author{Q. Vera Liao}
\affiliation{%
  \institution{Microsoft Research}
  \city{Montreal}
  \state{QC}
  \country{Canada}
}
\email{veraliao@microsoft.com}

\author{Alexandra Olteanu}
\affiliation{%
  \institution{Microsoft Research}
  \city{Montreal}
  \state{QC}
  \country{Canada}
}
\email{alexandra.olteanu@microsoft.com}

\author{Jina Suh}
\affiliation{%
  \institution{Microsoft Research}
  \city{Redmond}
  \state{WA}
  \country{USA}
}
\email{jinsuh@microsoft.com}

\author{Koustuv Saha}
\affiliation{%
  \institution{University of Illinois at Urbana-Champaign}
  \city{Urbana}
  \state{IL}
  \country{USA}
}

\email{koustuv.saha@gmail.com}

\renewcommand{\shortauthors}{Kawakami et al.}

\begin{abstract}
With the heightened digitization of the workplace, alongside the rise of remote and hybrid work prompted by the pandemic, there is growing corporate interest in using passive sensing technologies for workplace wellbeing. Existing research on these technologies often focus on understanding or improving interactions between an individual user and the technology. Workplace settings can, however, introduce a range of complexities that challenge the potential impact and in-practice desirability of wellbeing sensing technologies. Today, there is an inadequate empirical understanding of how everyday workers---including those who are impacted by, and impact the deployment of workplace technologies--envision its broader socio-ecological impacts. In this study, we conduct storyboard-driven interviews with 33 participants across three stakeholder groups: organizational governors, AI builders, and worker data subjects. Overall, our findings surface how workers envisioned wellbeing sensing technologies may lead to cascading impacts on their broader organizational culture, interpersonal relationships with colleagues, and individual day-to-day lives. Participants anticipated harms arising from ambiguity and misalignment around scaled notions of ``worker wellbeing,'' underlying technical limitations to workplace-situated sensing, and assumptions regarding how social structures and relationships may shape the impacts and use of these technologies. 
Based on our findings, we discuss implications for designing worker-centered data-driven wellbeing technologies.  
\end{abstract}




\maketitle

\section{Introduction}

With rapid transitions into a digital workplace, data-driven worker tracking technologies are increasingly considered ``commonplace'' across companies~\cite{yang2010tracking,muckell2017wearable}. Further motivated by the rise of remote and hybrid work prompted by the COVID-19 pandemic, organizations are beginning to explore the use of new data-driven technologies that aim to measure workers' wellbeing~\cite{zickuhr2021workplace,finnegan2022newNormal,nyt2022surveillance}. One form of worker wellbeing technologies growing in prominence includes the use of passive sensing methods~\cite{dasswain2020social,bleakley2022bridging,russell2021videoconferencing,yang2022distance}. Passive sensing methods are designed for unobtrusive collection and analysis of behavioral and contextual data from individuals in situ, for example, through mobile devices such as smartphones~\cite{cornet2018systematic}, wearables~\cite{rabbi2011passive}, and online and digital interactions~\cite{mitra2017spread,shami2015inferring}. Recent research has leveraged these technologies to infer a variety of physiological, psychological, and social wellbeing constructs~\cite{wang2014studentlife,wang2018sensing,rabbi2015mybehavior,binmorshed2019mood}, including workplace-specific measures~\cite{mirjafari2019differentiating,mark2016email,kucukozer2021designing}. Such approaches aim to address the limitations of traditional survey-based wellbeing measurements, which are comparatively harder to scale, may overlook temporal dynamics, and are susceptible to recall biases~\cite{fredrickson2000extracting,fisher1993social,tourangeau2000psychology}.

Much research and development efforts of workplace wellbeing technologies have been dedicated to understanding or improving interactions between an individual user and the technology. 
Workplace settings, however, introduce a range of complexities that may challenge the potential impact and desirability of such technologies. Indeed, a growing body of academic literature and industry reports have raised concerns around how profit-maximizing incentives and power dynamics could impact workers' use of workplace-governed technologies~\cite{pfeffer2020employers,cheon2021human,collective2022human,ajunwa2017limitless}. When deployed in a workplace setting, data-driven wellbeing technologies, in particular, may also risk lending themselves to heightened surveillance and privacy concerns from increasing workers' digital visibility, issues about social desirability~\cite{gianakos2002predictors}, 
amplified worker burdens around power asymmetries~\cite{introna2000workplace,miller1990governing}, coercive control~\cite{sewell2006coercion}, marginalization and biases~\cite{ruggs2013gone} and lack of explicit consent~\cite{hodson1994loyalty,chowdhary2023consent}---factors that also contribute to an ongoing problematic cultural shift towards commodified workers~\cite{littler1978understanding}. 

However, in the human-computer interaction (HCI), ubiquitous computing (UbiComp), and computer-supported cooperative work (CSCW) literature, where many such sensing innovations are blossoming on the technical front~\cite{mirjafari2019differentiating,binmorshed2019mood,mark2014capturing,schaule2018employing,binmorshed2022advancing}, there is an inadequate empirical understanding of how everyday workers envision passive sensing-based wellbeing technologies may live within a socio-ecological context. 
Like most other technologies, workplace wellbeing sensing technologies are often designed top-down (e.g., the notion of ``worker wellbeing'' is conceptualized and operationalized primarily through the decisions made by researchers and product teams~\cite{mohr2017three,mohr2017accelerating}). With growing corporate interest in adopting these technologies in practice, it is critical to center worker perceptions around the desirability and anticipated (mis)use of these technologies~\cite{park2022designing,kaur2022didn,adler2022burnout}. Without centering the voices of workers, most of whom typically have little to no decision latitude for deploying these technologies, the technologies might eventually fail post-deployment, producing individual and societal harms along the way~\cite{grudin1988cscw}.

Progressing a more holistic understanding of the socio-ecological impacts that workers envision---including perceptions around \textit{why} certain harms may be more likely to arise in the workplace, and how they could be addressed and mitigated, if at all---could help more proactively inform growing trends in the in-practice adoption of data-driven worker wellbeing technologies. Workers who are subject to these technologies, and who often do not have a say in informing their design or deployment, are at the highest risk of being impacted in undesirable ways. At the same time, it is just as critical to understanding the potential motivations and envisioned impacts of technology builders and organizational governors who may drive decisions surrounding their deployment.  
Workplace technologies implicate a broad range of stakeholders, each varying in the position and power they hold within an organization. Such differences among stakeholders may give rise to differing incentives, and, therefore, diverse perceptions of the benefits and harms of these technologies.

Our study aims to develop a systematic \textbf{understanding of multiple stakeholder perspectives on the envisioned beneficial and harmful impacts of wellbeing sensing technologies deployed in the workplace}. We focus on how the stakeholders perceive wellbeing technologies may interface with the complex organizational structures, interpersonal dynamics, and individual preferences and needs within the workplace.
We conducted semi-structured interviews with 33 participants from three broad stakeholder groups, 1) organizational governors---who may impact decisions to deploy these technologies, 2) AI builders---who are involved with the design, research, and development of these technologies, and 3) worker data subjects---who would primarily be impacted by these technologies. Through the lens of how power might operate within an organization, we \textit{studied up}~\cite{barabas2020studying} the organizational governors, alongside the AI builders and worker data subjects on the ground. 
To guide our interviews, we sketched and used four storyboards inspired by prior work (e.g., ~\cite{huang2019digital,xie2018work,mark2016email,DasSwain2019FitRoutine,saha2019libra}) to depict different worker wellbeing sensing innovations and settings and to elicit participants' reactions towards different ways wellbeing sensing technologies may be used in the workplace.

By disentangling the harms and benefits workers envisioned across workplace-specific levels inspired by the socio-ecological model~\cite{catalano1979health}, we surface how data-driven worker wellbeing technologies may impact, and are subject to being impacted by, organizational culture and structures, interpersonal relations, and individual needs and desires. \edit{Overall, our findings clarify how and why common assumptions shaping the design and use of workplace wellbeing technologies may later impact \textit{who}--if not the intended end-user worker--benefits from these technologies.} Envisioning the potential uses of these technologies, participants expressed concerns around how power imbalances and workplace incentives may lead their organization or manager to repurpose wellbeing technologies in ways that deprioritize, or actively detriment, their own wellbeing. Reflecting on the design of these technologies, participants' responses also surfaced ambiguity and misalignment around the notions of ``workplace wellbeing'' that resonate with individual workers and the measures for ``workplace wellbeing'' that are typically defined and scaled.

\edit{Drawing on our findings, we discuss how participants contested notions of ``workplace wellbeing,'' as differing across individuals and with proxies used by the technology. We discuss implications for how these findings challenge the efficacy and desirability of properties typically celebrated in sensing approaches, like scalability and objectiveness. Moreover, we extend calls for HCI researchers to explore questions beyond technology design by discussing opportunities for additionally supporting the design of worker-centered \textit{policies} and avenues for technology refusal. }Finally, we conclude with design recommendations for organizations interested in developing wellbeing technologies and researchers designing or studying these technologies. 

\section{Background and Related Work}\label{sec:related}
\subsection{Measuring Workplace Wellbeing}

\edit{Wellbeing is a heavily used but complex construct. It is a combination of multiple factors-–-crosscutting people's personal, workplace, cultural, and other experiences.  
``Wellbeing,'' by definition extends beyond traditional definitions of health, and higher wellbeing indicates that in some sense the condition of an individual or a group is positive and they are happy and productive~\cite{schulte2010well}. When considering workplace, International Labour Organization defines it as relating to all aspects of working life, from the quality and safety of the physical environment, to how workers feel about their work, their working environment, the climate at work, and work organization~\cite{ILO2023}.~\citeauthor{hickok2022framework} outlines six pillars of worker wellbeing, 1) human rights, 2) physical safety and health, 3) financial wellbeing, 4) intellectual wellbeing, 5) emotional wellbeing, and 6) purpose and meaning.} 

Research notes that the wellbeing of individuals at the workplace also translates to individual, collective, and organizational success~\cite{knight2006impact}. A rich body of literature in organizational psychology and organizational behaviors has extensively studied the causes and correlates of improving wellbeing and performance at workplaces~\cite{biggio2013well,bryson2014does}. Research postulates that workplace wellbeing is associated with the interaction between individual characteristics and those of the working and organizational environment~\cite{biggio2013well}. A body of research has emphasized and studied the challenges of workplace stress---stress that arises if the demands of an individual's role and responsibilities exceed their capacity and capability to cope~\cite{colligan2006workplace}. 
To understand workplace stress better,~\citeauthor{de2013objective} proposed the importance of considering subjective wellbeing in the workplace as a coarse construct that leads to objective benefits across the major life domains of 1) health and longevity, 2) income, productivity, and organizational benefits, and 3) individual and social behaviors~\cite{de2013objective}.

While wellbeing measurements are most validated through survey methods, research also cautions that survey questions may be responded to ``carelessly''~\cite{kam2015careless}, and---because of their underlying parsimonious design---survey questions may not be interpreted the same way across individuals, and the responses are often reliant on an individual's retrospective recall; which may bias the results~\cite{fredrickson2000extracting,fisher1993social,van2008faking}.
Surveys are also challenging to do at the population-level scale and can often not capture high-granular temporal phenomena like wellbeing~\cite{andrade2020limitations}. 
Further, Pew estimates that survey response rates tend to be very low (9\% in the U.S.)~\cite{baldauf1999examining,baruch2008survey}.
To mitigate some of the confounds of static long-form surveys, research has proposed ecological momentary assessments (EMAs) (also dubbed experience sampling)~\cite{scollon2009experience,stone1998comparison}, where participants are prompted to respond to short survey items in-the-moment~\cite{stone1998comparison}. 
HCI and UbiComp research has extensively studied and used EMAs (or active sensing, when integrated with mobile-sensing technologies), including in the workplace settings~\cite{howe2022design,mattingly2019tesserae,binmorshed2019mood,kucukozer2021designing}. However, these methods do require active engagement and induce a response bias on participants through disruptions~\cite{suh2016developing}. This entails the requirement to balance between the construct validity of distributing short survey items with participant compliance~\cite{chan2018students}, and excessive use of EMAs can be challenging in longitudinal and robust data collection~\cite{froehlich2007myexperience}. 

Consequently, more passive and unobtrusive sensing have been emerging to collect and infer wellbeing constructs~\cite{cornet2018systematic,rabbi2011passive}. 
The ubiquity and widespread use of smartphones and wearables have enabled the collection of longitudinal and dense human behavioral cues at scale~\cite{wang2014studentlife,wang2018sensing}. Within the context of workplace settings, prior work has studied the use of wearables in understanding aspects of job satisfaction~\cite{olguin2011sensor}.~\citeauthor{mark2016email} studied how email interactions associate with workplace stress and productivity~\cite{mark2016email}. Another work~\cite{mark2014bored} leveraged digital activities of information workers to understand work patterns and attentional states~\cite{mark2014bored}. In addition, prior research has used multimodal sensing through smartphone, wearable, bluetooth, and wireless sensors to try to infer job performance~\cite{mirjafari2019differentiating}, mood and cognition~\cite{binmorshed2019mood,mark2014capturing,schaule2018employing,binmorshed2022advancing,robles2021jointly}, social interactions~\cite{brown2014architecture,matic2014mobile}, organizational personas~\cite{dasswain2019multisensor}, organizational fit~\cite{DasSwain2019FitRoutine}. Further, with the widespread use of online and social media technologies, prior research has argued that such data can serve as longitudinal, historical, and verbal passive sensors~\cite {saha2019tessSocial}. 
A series of research with IBM's Beehive platforms have studied how people's social interactions associate with various workplace dynamics~\cite{dimicco2008motivations,dimicco2009people,farzan2008results,geyer2008use}. Other prior research has used such online and social media data in studying various dimensions and correlates of workplace wellbeing, such as engagement~\cite{mitra2017spread,shami2015inferring}, mood and affect~\cite{de2013understanding,shami2014social}, organizational relationships~\cite{brzozowski2009watercooler,gilbert2012phrases,mitra2012have}, organizational culture~\cite{dasswain2020culture,xu2016predicting}, and job satisfaction~\cite{saha2021job,lee2017study}. Together, this body of research has highlighted the promise of passive sensing technologies in measuring and understanding the workplace wellbeing dynamics of individuals. 

Despite the promise, these measurements have primarily been evaluated in research and observational settings, where subjects are removed from the multiplex, incentive structures, and dynamics in real-world workplace settings. There is a lack of understanding of the in-practice utility and ecological validity of these approaches. Research has revealed other limitations of these sensing-based data-driven wellbeing measurements, including how poor proxies could lead to fairness and bias issues~\cite{jacobs2021measurement,olteanu2019social,passi2019problem}. There are possible mismatches in people's self-perceived versus passively sensed and automatically inferred wellbeing~\cite{das2022semantic,roemmich2021data}.~\citeauthor{kaur2022didn} found that the misalignments were associated with the lack of workplace activity and context data that the sensing technology might not comprehensively include, and the authors questioned the construct validity of the measurements~\cite{kaur2022didn}.~\citeauthor{veale2021demystifying} noted that ``those claiming to detect emotion use oversimplified, questionable taxonomies; incorrectly assume universality across cultures and contexts.'' This body of research critically questions the applicability of digital technologies to automatically infer wellbeing~\cite{docherty2022re,howell2018emotional,corvite2022data,dasswain2023algorithmic,constantinides2022good}.
\edit{In parallel, the multiple transitions of workplaces in recent years, including the growth of remote and hybrid work~\cite{yang2022effects}, also amplified by the pandemic, have seen a proliferation of digital traces such as video conferencing~\cite{bleakley2022bridging,russell2021videoconferencing} and remote collaboration tools~\cite{yang2022distance,haliburton2021charting}---the data from which can also be passively collected to measure workplace behaviors.}

There is growing excitement among companies about adopting these passive technologies to support wellbeing~\cite{litchfield2021workplace}. \textit{But, are we there yet?} Our work draws motivation from~\citeauthor{mohr2017three} in contributing empirical insights about filling the gap between translating research to practice with respect to worker wellbeing sensing technologies. Drawing on the social-ecological model~\cite{catalano1979health}, that an individual worker is situated within their social ecology, we study the impacts of worker wellbeing technologies at individual, interpersonal, and organizational levels. We conduct our study through storyboard-design-based interviews, where our storyboards are theoretically driven from the literature~\cite{huang2019digital,xie2018work,mark2016email,DasSwain2019FitRoutine,saha2019libra,saha2023focus,dasswain2023focused}, and we examine the participants' perceptions of such envisioned technologies in the workplace.

\subsection{Data-Driven Worker Activity Technologies in Organizational Settings}
The deployment and use of workplace technologies are situated amidst the complexities introduced by the very nature of workplace settings and socio-organizational factors. Individual workers often have limited power within organizations, and rely on organizations for not only their livelihoods but also for health insurance and other essential benefits. Several scholars have raised concerns about the harms workers can face due to their powerlessness, especially when AI-based technologies are introduced in the workplace~\cite{cheon2021human,collective2022human,crawford2016ai,eubanks2018automating}.
A major area of concern with AI- and sensing-based technologies is workplace surveillance and intrusions into workers' privacy~\cite{ball2010workplace,lane2003naked, nolan2003privacy, lyon2003surveillance, rosenblat2014workplace, moore2017regulating,watkins2007workplace}. Prior work highlights that electronic surveillance in the workplace can have multiple justifications including surveillance as coercive control and surveillance as caring~\cite{watkins2007workplace,sewell2006coercion}. ~\citeauthor{watkins2007workplace} found that workers expected explicit boundaries that employers should not cross with respect to surveillance~\cite{watkins2007workplace}. Further, with the rise of new web and social media technologies,~\citeauthor{ghoshray2013employer} noted that employer surveillance shares a competing relationship with a worker's subjective expectation of privacy, and can influence a worker's use of these platforms~\cite{ghoshray2013employer}. Critical legal and computing scholars have situated these technologies within the history of workplace surveillance as advances that enable ``limitless worker surveillance''~\cite{ajunwa2017limitless}. These technologies have facilitated easier and more extensive means of workplace surveillance and lower costs~\cite{ajunwa2017limitless}.
Surveillance-related concerns are further exacerbated by the ``black box'' nature of workplace technologies --- the lack of transparency further encroaches on workers' autonomy and personhood and stymies their agency and ability to resist~\cite{ajunwa2020black}. This lack of transparency and the resulting effects are especially concerning given the invasive and intimate nature of the data that is often collected by sensing technologies~\cite{bodie2017law}.

The intimate nature of the collected data, as facilitated by sensing technologies, introduces another concern with these technologies: the potential for punishment or discrimination. Critical scholars argue that the primary focus of these technologies is to ``capture and control attention''~\cite{till2019creating} and, by conflating health with productivity, these could be another mechanism for organizations to exert what has been dubbed as bio-power over workers~\cite{mantello2021bosses}. In an early work,~\citeauthor{foucault1979discipline} conceptualized ``disciplinary'' power, as the power ``watchers'' have to observe, scrutinize, and control the behavior of individuals within the institutions they control. Bio-power is an extension of ``disciplinary'' power where the omnipresence of the ``watchers'' leads the ``watched'' to begin to internalize and self-enforce the existing rules and norms (in this case, ideals of productivity that became codified into these technologies), and start behaving in the required manner without coercion~\cite{foucault1979discipline}. 
Recent research has emphasized that the design of these technologies, especially the passive and constantly-present nature, and individualistic focus, can further amplify bio-power~\cite{cecchinato2021self-tracking, manokha2020implications, moore2016quantified}. Such power could cause disproportionate impacts on workers 
from disadvantaged populations~\cite{bodie2017law, mantello2021bosses, stark2020don}.

In the area of supporting worker wellbeing, a recent interview study with workers across diverse, purposely sampled, progressive corporations, revealed that companies ``often viewed keeping employees healthy primarily as a means to profitability rather than an end in itself and rationalized stressful workplaces as necessary and non-changeable''~\cite{pfeffer2020employers}.
These profit-motivated incentives are also seen in the motivations behind corporate wellness programs~\cite{ajunwa2017limitless,ajunwa2016health}, which may also include passive sensing technologies~\cite{chung2017finding}.
The imbalance in whose interests are often served by these technologies reflects what \citeauthor{maltseva2020wearables} describes as one of the key challenges with implementing passive sensors in the workplace: the colonization of the private domain by the professional domain, by turning employees into resources to be extracted~\cite{maltseva2020wearables}. 

Given the complexities surrounding power imbalances and resulting harms, research has discussed the ethics of designing and deploying workplace technologies~\cite{moradi2020future,posada2020future,roossien2021ethics}. Prior work proposed participatory, value-sensitive, and worker-centered design methodologies to mitigate technological harms~\cite{zhang2022algorithmic,lee2021participatory, hickok2022framework,liao2019enabling}. 
\citeauthor{madaio2020co} adopted a participatory approach to co-design checklists of organizational challenges in AI ethics~\cite{madaio2020co}, and~\citeauthor{zhu2018value} proposed Value Sensitive Algorithm Design~\cite{zhu2018value} in early stages of algorithm creation to avoid compromising stakeholder values~\cite{zhu2018value}. Given the socio-organizational situatedness of these technologies, research recognizes the detrimental effect of a techno-centric view on AI technologies~\cite{shneiderman2020human,sabanovic2010robots}. If individuals do not realize the values or use of these systems, there is a lower likelihood of their acceptability into the individual and organizational workflows~\cite{makarius2020rising,wolf2019evaluating}. 

Building on the above body of research, our work aims at shedding light on the perceived needs, concerns, and desires of workers regarding wellbeing sensing in the workplace. Adopting a worker-centered research design, our work helps to uncover previously understudied aspects of how different system actors, including more powerful system actors who have the ability to shape the systems and whose interests drive the development of these systems. Our work is inspired by the anthropological concept of ``studying up''~\cite{barabas2020studying}, or the process of examining the most powerful to understand how a system truly works and all of the ways it disadvantages the less powerful. Our work contributes to and bears implications in accounting for multi-stakeholder voices in designing, deploying, and decision-making with respect to worker wellbeing technologies in the workplace.

\subsection{Foreseeing Harms of AI and Sensing-Based Technologies}
While AI and computing have proliferated significantly in the last few decades and been more and more embedded in our lives, they have also caused harm at both individual and societal levels~\cite{wagner2021measuring}.~\citeauthor{boyd2012critical} noted that data-driven technologies that leverage the vividly available digital data of individuals can reinforce the troubling perception of these technologies as ``Big Brother, enabling invasions of privacy, decreased civil freedoms, and increased state and corporate control''~\cite{boyd2012critical}. Recently,~\citeauthor{raji2022fallacy} highlighted how AI systems that are being deployed often do not work as intended~\cite{raji2022fallacy}. Although anticipating all unintended uses and consequences of AI has proven difficult~\cite{boyarskaya2020overcoming,coston2022validity,raji2020closing}, there have been growing efforts to map and taxonomize possible failures, risks, and harms, often in an attempt to gain a more comprehensive understanding of the issues that could arise within and across different AI deployment and use settings~\cite{raji2022fallacy,amershi2019guidelines,chancellor2019taxonomy,barocas2017problem,crawford2017trouble,ehsan2021expanding,holstein2019improving}. These include, taxonomy of AI failures~\cite{raji2020closing}, guidelines for designing human-AI interaction technologies~\cite{amershi2019guidelines}, taxonomy of ethical tensions between AI and practitioners~\cite{chancellor2019taxonomy}, taxonomies of roles of AI consumers~\cite{arrieta2020explainable,tomsett2018interpretable}, question-bank of explainable AI~\cite{liao2020questioning}, and guidelines for minimizing risks with automated emotion recognition technologies~\cite{hernandez2021guidelines}. Relatedly, recent research has seen datasheets for datasets~\cite{gebru2021datasheets}, explainability fact sheets~\cite{sokol2020explainability}, and model cards~\cite{mitchell2019model} towards building responsible and ethical AI technologies.  
Further, there is a growing recognition that the priorities of different stakeholders are shaped by their backgrounds and personal experiences~\cite{jakesh2021how}, and that more participatory approaches are required to understand the range of needs and concerns of different stakeholders and practitioners~\cite{madaio2022assessing,coston2022validity,wagner2021measuring}.

Relatedly, research has investigated algorithmic impacts in different workplace settings and worker populations, including ridesharing~\cite{lee2015working,cram2020algorithmic}.~\citeauthor{lee2021participatory} adopted participatory design to study how worker wellbeing can be incorporated into algorithmic management so that the workplace can be better optimized for workers~\cite{lee2021participatory}.~\citeauthor{hickok2022framework} proposed a framework for worker wellbeing in an AI-integrated workplace, including six pillars of worker wellbeing across 1) human rights, 2) physical safety and health, 3) financial wellbeing, 4) intellectual wellbeing, 5) emotional wellbeing, and 6) purpose and meaning~\cite{hickok2022framework}. Recently,~\cite{park2022designing} studied designing fair worker-centered AI for Human Resource Management~\cite{park2022designing}, and~\citeauthor{adler2022burnout} studied the tensions between how to measure workplace wellbeing better and how these technologies remake the boundaries of information flows between worker and workplace~\cite{adler2022burnout}.

Our work draws inspiration from and contributes to the literature investigating the ethical challenges and concerns about possible adverse impacts that the use of AI technologies can give rise to~\cite{raji2022fallacy,crawford2017trouble,barocas2017problem}. A major contribution of our work involves foreseeing the harms of workplace wellbeing technologies and revealing their utility and nuances. Our problem is situated in this space of deploying sensing technologies for worker wellbeing, which is envisioned and believed to be theoretically backed; however, there are currently only a few recent attempts at large-scale real-world deployments. Therefore, we consider time-critical to empirically study and surface workers' potential concerns and needs. Our work motivates us to revisit and re-imagine these technologies' very existence and design, and what factors and mitigation strategies to consider with respect to these technologies.

\section{Study Design and Methods}  
To examine different organizational stakeholders' concerns about and desire to use sensing-based wellbeing technologies in the workplace, we conducted semi-structured interviews with 33 participants across three stakeholder groups, broadly construed as: organizational governors, AI builders, and workers (i.e. data subjects and/or target users of wellbeing sensing technologies). Participants were recruited to represent a broad range of stakeholders who would impact or be impacted by the design and deployment of workplace wellbeing technologies. Before conducting the interviews, we sketched four storyboards, each depicting different ways in which a hypothetical wellbeing sensing technology can be used in a workplace setting. During the interviews, we used the storyboards 
as a provocative tool to elicit participants’ reactions toward integrating wellbeing sensing technologies into a workplace context. In the following, we describe the recruitment process and participant groupings, the design of the study protocol and storyboards, and our analysis method. 

\subsection{Participants and Recruitment} 
We aimed to recruit participants across a range of roles with varying positions and power around implementing and using wellbeing sensing technologies in the workplace. Our study is scoped to focus on information workers~\cite{mark2014bored,mattingly2019tesserae}. Information workers process and work with information rather than physical objects and, in recent years, do so typically using computing technologies. 
We recruited participants from 12 organizations of varying sizes, types (technology, consulting, research-based, non-profit, etc.), and four global locations (U.S., U.K., Canada, and India). \edit{The countries included in the study were selected to work around access restrictions placed by our institution's IRB, as well as privacy and regulatory considerations.}

Our recruitment was carried out in three weeks in June and July 2022. We adopted a snowballing strategy by posting on organizational email lists and directly contacting information workers in different roles through LinkedIn direct messaging and emails. \edit{We also reached out to our (different coauthors') contacts at multiple companies to inform them about our study and request that they share the interest forms with their contacts and relevant mailing lists.} 
Interested respondents were asked to complete an interest form that asked for information on their role, workplace tenure, organizational size, familiarity with AI, ML, or data science, and if they have served in a leadership or managerial position. Out of the 198 responses that we received, we invited a subset of participants to maximize diversity and balance across answers to the above questions. This led to a final subset of 33 participants that consented to participate in the study, and we interviewed. 
Each participant was compensated with Amazon gift vouchers of \$50 USD (or equivalent amounts in local currencies).

We group these participants into three stakeholder groups, 1) \textit{Organizational Governors (\textsf{G})}: individuals in leadership, human resources, policy-making, and corporate legal and governance roles, who are primarily responsible for decision-making surrounding the development, deployment, and use of worker technologies, 2) \textit{Builders (\textsf{A})}: individuals associated with building AI and sensing technologies, in the roles of design, engineering, research, and product management, 3) \textit{Worker Data Subjects (\textsf{W})}: workers who would be subjected to the use of workplace wellbeing technologies, including people serving both management (people, product, general, etc.) and rank-and-file roles. 
Our participant pool included an equal number (11) of participants from each group of stakeholders. \autoref{table:participant_demographics} and \autoref{table:participant_details} provide descriptive summaries about our participant pool. We note that individuals in the first two groups would likely serve multiple roles in an organization (e.g., a builder is also a worker data subject). We accounted for this in our study protocol by clarifying that participants can switch hats anytime when discussing the storyboards.

\begin{table}[t]
\sffamily
\caption{\edit{Distribution of participants in the study, along with count and percentage of participants in the pool. The data is aggregated to protect participant privacy.}}
\vspace{-0.5em}
\label{table:participant_demographics}
  \centering
    \footnotesize
\begin{minipage}[t]{0.45\columnwidth}
    \begin{tabular}{lrr}
 \textbf{Response} & \textbf{Count} & \textbf{Percentage} \\
 \cmidrule(lr){1-1}\cmidrule(lr){2-2}\cmidrule(lr){3-3}

\rowcollight \multicolumn{3}{l}{{{\textbf{Gender}}}}\\
Woman & 16 & 48.48\% \\
Man & 15 & 45.45\%\\
Non-binary/gender diverse & 1 & 3.03\%\\
Prefer not to answer & 1 & 3.03\%\\
\hdashline
\rowcollight \multicolumn{3}{l}{{{\textbf{Race}}}}\\
White/Caucasian & 17 & 51.52\%\\
Asian & 11 & 33.33\%\\
Prefer not to answer & 5 & 15.15\%\\
\hdashline
\rowcollight \multicolumn{3}{l}{{{\textbf{Age}}}}\\
18-24 & 9 & 27.27\%\\
    25-34 & 7 & 21.21\%\\
    35-44 & 6 & 18.18\%\\
    45-54 & 9 & 27.27\%\\
    65+ & 1 & 3.03\%\\
    Prefer not to answer & 1 & 3.03\%\\
\hdashline
\rowcollight \multicolumn{3}{l}{\textbf{Country}}\\
U.S. & 21 & 63.64\%\\
    Canada & 5 & 15.15\%\\
    U.K. & 3 & 9.09\%\\
    India & 4 & 12.12\%\\
\hdashline
\rowcollight \multicolumn{3}{l}{\textbf{Tenure at current company (years)}}\\
<1 Y. & 11 & 33.33\%\\
1-3 Ys. & 5 & 15.15\%\\
3-5 Ys. & 3 & 9.09\%\\
 5-7 Ys. & 5 & 15.15\%\\
 >7 Ys. & 6 & 18.18\%\\
\bottomrule
\end{tabular}
\end{minipage}\hfill
\begin{minipage}[t]{0.52\columnwidth}
\begin{tabular}{lrr}
\textbf{Response} & \textbf{Count} & \textbf{Percentage} \\
 \cmidrule(lr){1-1}\cmidrule(lr){2-2}\cmidrule(lr){3-3}
\rowcollight \multicolumn{3}{l}{\textbf{Organization's Employee Size}}\\
<50 & 2 & 6.06\%\\
50-1000 & 1 & 3.03\%\\
1K-10K & 1 & 3.03\%\\
>10K & 29 & 87.88\%\\
\hdashline
 \rowcollight \multicolumn{3}{l}{\textbf{What best describes your profession?}}\\
Human Resources & 4 & 12.12\%\\
    Design \& Creative & 1 & 3.03\%\\
    Engineering \& Software Dev. & 3 & 9.09\%\\
    Management & 8 & 24.24\%\\
    Research & 8 & 24.24\%\\
    Sales and Marketing & 4 & 12.12\%\\
    Legal & 3 & 9.09\%\\
    Business Operations & 1 & 3.03\%\\
    Business Development & 1 & 3.03\%\\
\hdashline
\rowcollight \multicolumn{3}{p{0.9\columnwidth}}{\textbf{Do you work on topics related to AI, ML, or data science?}}\\
Yes & 16 & 48.48\% \\
    No & 13 & 39.39\%\\
    Other & 2 & 6.06\%\\
\hdashline
\rowcollight \multicolumn{3}{p{0.9\columnwidth}}{\textbf{Do you currently or in the past served in a leadership or managerial position?}}\\
Yes & 15 & 45.45\%\\
    No & 17 & 54.52\%\\
    Blank & 1 & 3.03\%\\
    \\\\
\bottomrule
\end{tabular}
\end{minipage}

\end{table}

\begin{table}[t]
\sffamily
\caption{Distribution of participants in the study across various stakeholder groups.}
\vspace{-1.2em}
\label{table:participant_details}
\begin{minipage}[t]{0.25\columnwidth}
  \centering
    \footnotesize
    \begin{tabular}{ll}
\setlength{\tabcolsep}{1pt}\\
 \multicolumn{2}{l}{\textbf{Organizational Governors}}\\
 \toprule
\rowcollight \textbf{ID}&\textbf{Role / Discipline}\\
G1 & HR - Talent Sourcing\\
G2 & HR   \\
G3 & HR Legal \\
\hdashline
G4 & Legal \\
G5 & Chief Business Officer \\
G6 & AI Policy-Maker  \\
\hdashline
G7 & AI Ethics Director \\
G8 & HR \\
G9 & HR Director  \\
\hdashline
G10 & Program Manager \\
G11 & Privacy Architect \\
\bottomrule
\end{tabular}
\end{minipage}\hfill
\begin{minipage}[t]{0.29\columnwidth}
  \centering
    \footnotesize
    \begin{tabular}{ll}
\setlength{\tabcolsep}{1pt}\\
\multicolumn{2}{c}{\textbf{Builders}}\\
\toprule
\rowcollight \textbf{ID}&\textbf{Role / Discipline}\\
A1 & Research \\
A2 & Engineering \\
A3 & Engineering \& Development\\
\hdashline
A4 & Design \\
A5 & Research \\
A6 & Research \\
\hdashline
A7 & Research \\
A8 & Design and Research \\
A9 & Research and Engineering\\
\hdashline
A10 & Research\\
A11 & Product Management\\
\bottomrule
\end{tabular}
\end{minipage}\hfill
\begin{minipage}[t]{0.39\columnwidth}
  \centering
    \footnotesize
    \begin{tabular}{ll}
	\setlength{\tabcolsep}{1pt}\\
\multicolumn{2}{c}{\textbf{Worker Data Subjects}}\\
\toprule
\rowcollight \textbf{ID}&\textbf{Role / Discipline}\\
W1 & General Management\\
W2 & Compliance Management\\
W3 & Project Management\\
\hdashline
W4 & Project Management\\
W5 & Business Development\\
W6 & Marketing Consultancy\\
\hdashline
W7 & Sales\\
W8 & Sales \\
W9 & Business Operations\\
\hdashline
W10 & Financial Relationship Management\\
W11 & Strategy Consultancy\\
\bottomrule
\end{tabular}
\end{minipage}
\end{table}

\subsection{Study Design} 
With the rise of media coverage on workplace tracking technologies~\cite{Barbaro2022,Kantor2022,Leonhardt2022,KANTOR2022b,Hernandez2020,Finnegan2022}, the idea of collecting worker data to measure wellbeing may prompt participants to imagine very different technologies. Additionally, passive sensing may be an unfamiliar method to a participant. Therefore, asking participants about their concerns or desires towards envisioned wellbeing sensing technologies may lead to responses that are not grounded in realistic scenarios. To establish common ground across participants, we leveraged pre-defined storyboards to elicit stakeholder perspectives. 

Storyboards are used to solicit people's immediate reactions and underlying needs towards a wide range of hypothetical scenarios depicted in storyboard sketches~\cite{davidoff2007rapidly,holstein2019designing}.
We leverage storyboards to surface participants' boundaries and latent needs towards different forms of wellbeing sensing technologies. 
Using storyboards also allows the introduction of other organizational stakeholders into the storyboard images. This creates an opportunity for stakeholders to react honestly to how they would feel about their colleagues or managers using these technologies on or with them. Such an honest reaction would be challenging to elicit in an actual workplace involving multiple stakeholders amidst possible power dynamics influencing participant responses.

We designed and incorporated four storyboards in our study that cover a range of wellbeing sensing technologies proposed in the research literature.
For each storyboard, we aimed to go more in-depth into understanding the underlying \textit{why} behind participants' immediate reactions. 
We first describe how we incorporate storyboards in our interview protocol, followed by how we constructed the storyboards.

\begin{figure}
\centering
    \begin{subfigure}[b]{0.5\columnwidth}
    \centering
    \includegraphics[width=\columnwidth]{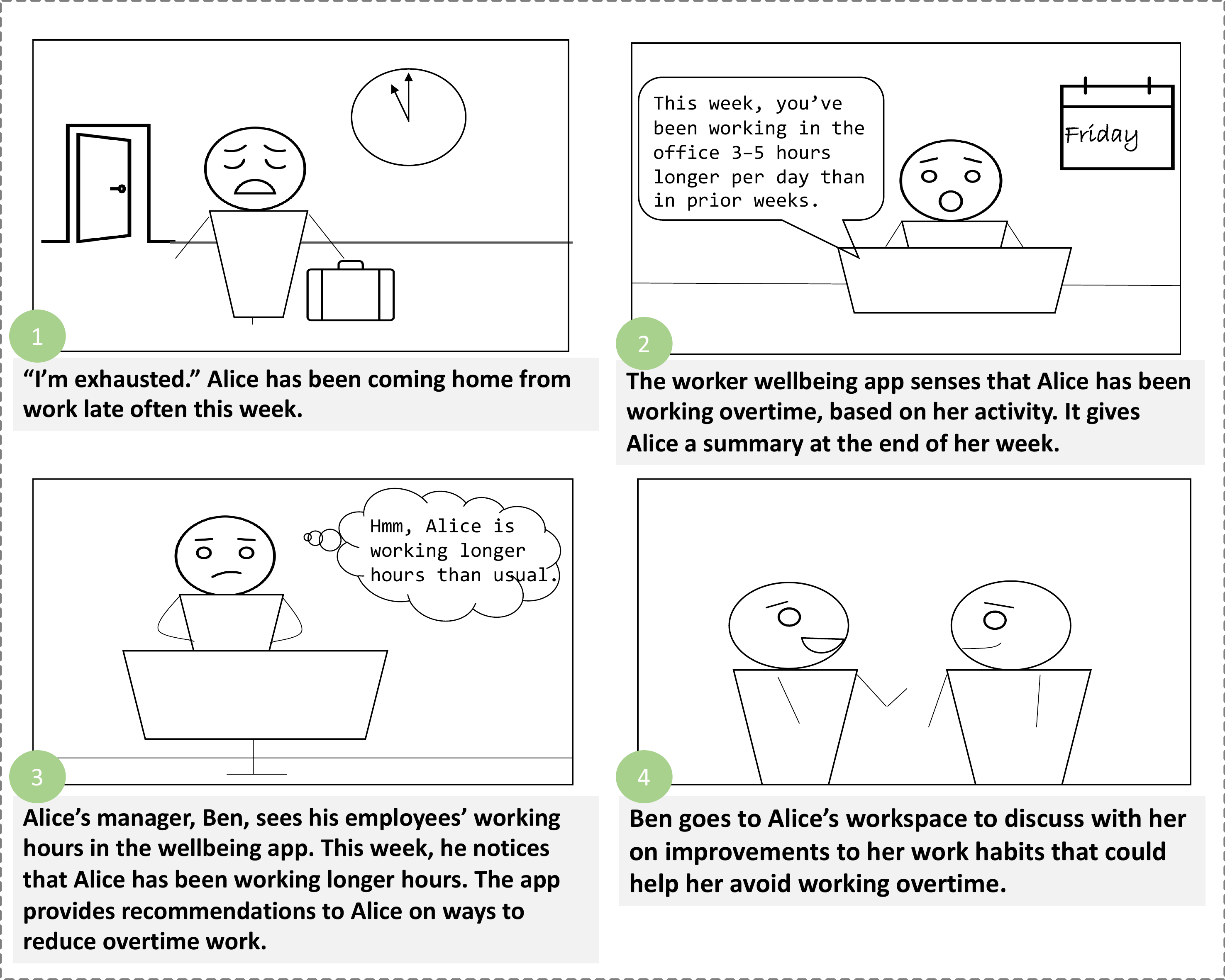}
    \caption{Storyboard \SBWPS{}: Work Patterns and Schedules}
    \label{fig:SB1}
    \end{subfigure}\hfill
    \begin{subfigure}[b]{0.5\columnwidth}
    \centering
    \includegraphics[width=\columnwidth]{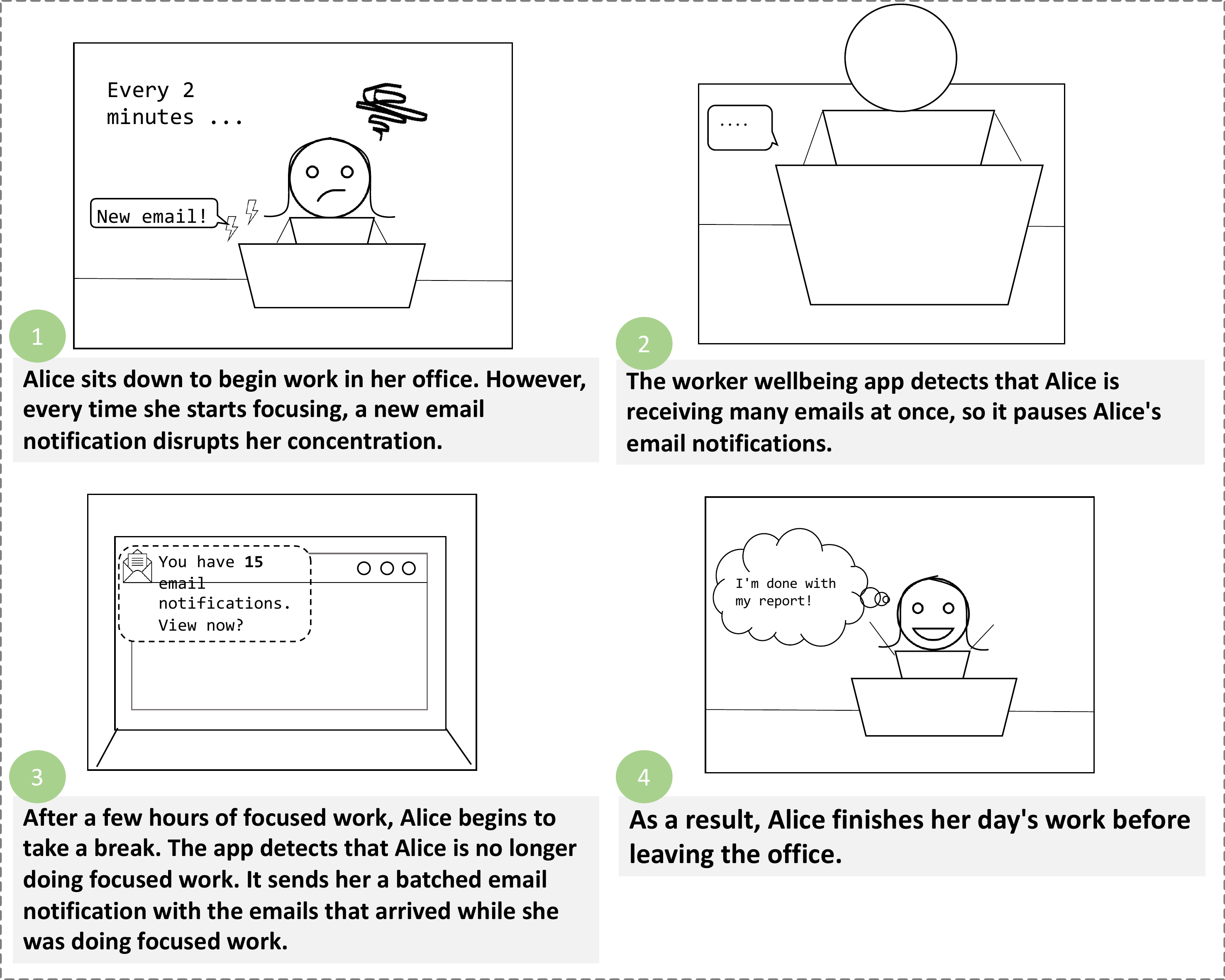}
    \caption{Storyboard \SBNFT{}: Email Notifications \& Focus Work}
    \label{fig:SB2}
    \end{subfigure}\hfill
    \begin{subfigure}[b]{0.5\columnwidth}
    \centering
    \includegraphics[width=\columnwidth]{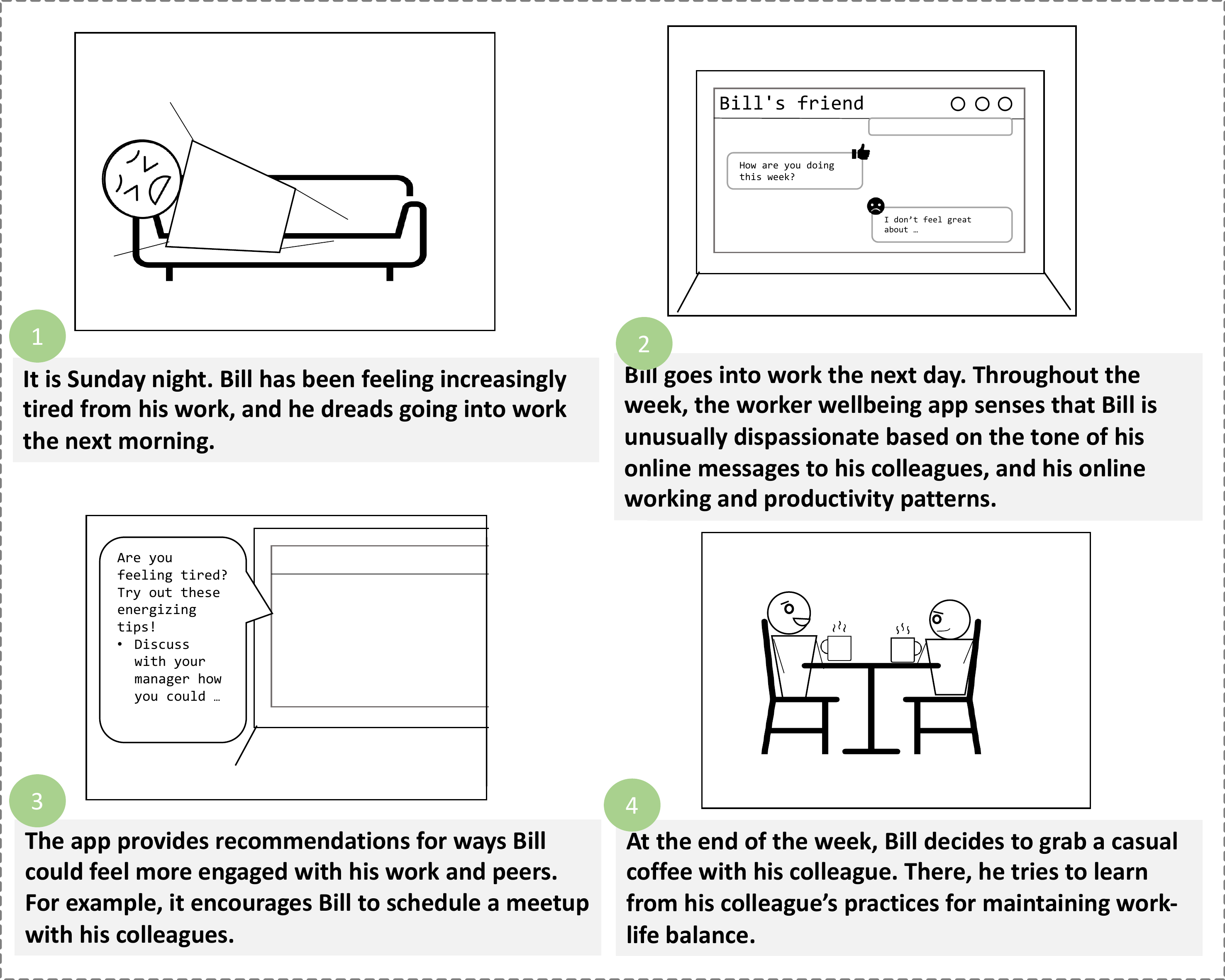}
    \caption{Storyboard \SBSIE{}: Emotions and Social Interactions}
    \label{fig:SB3}
    \end{subfigure}\hfill
    \begin{subfigure}[b]{0.5\columnwidth}
    \centering
    \includegraphics[width=\columnwidth]{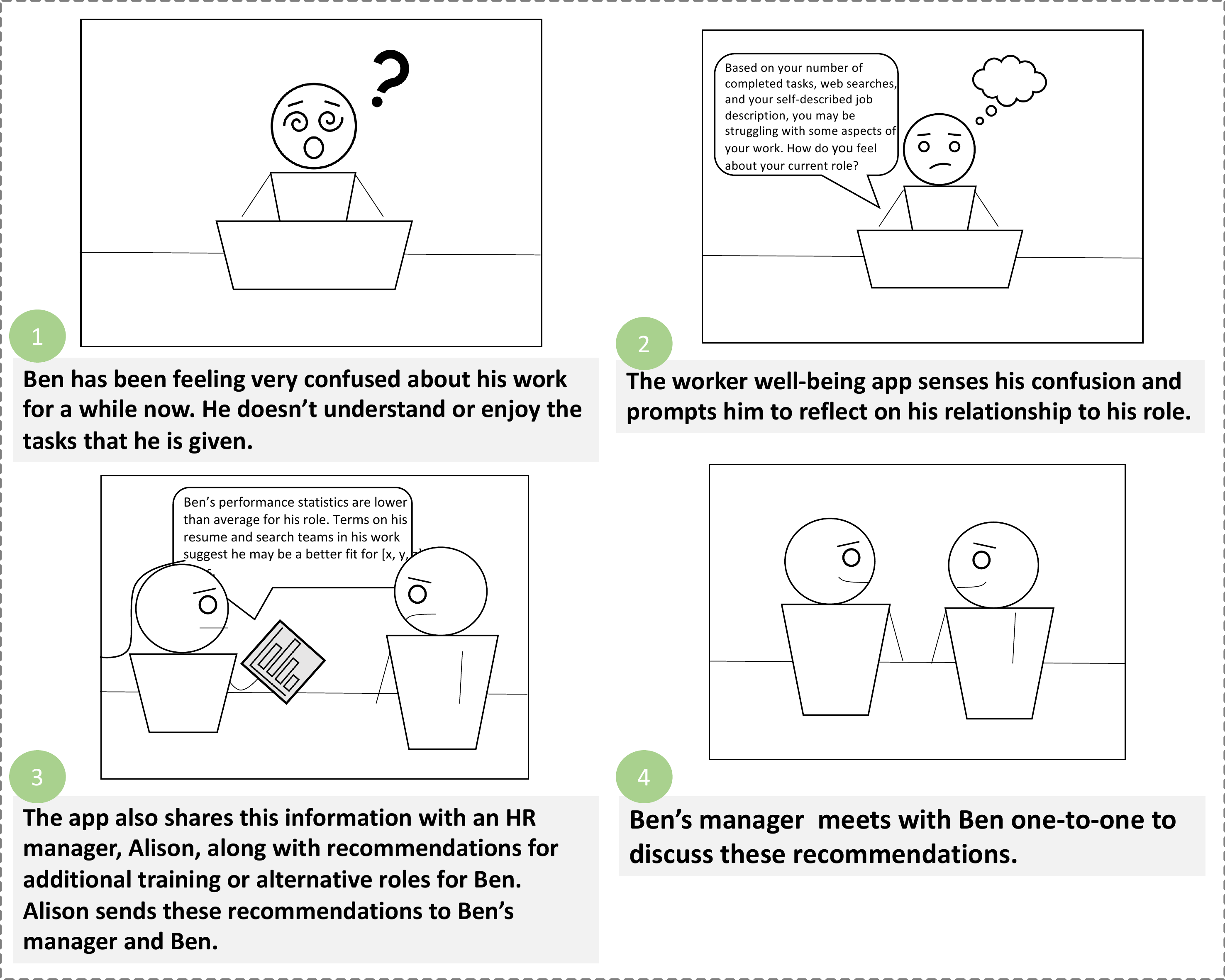}
    \caption{Storyboard \SBJFP{}: Productivity and Job Fit}
    \label{fig:SB4}
    \end{subfigure}\hfill
    \caption{Storyboards used to elicit stakeholder attitudes towards worker wellbeing sensing technologies. These storyboards are theoretically-driven by a range of prior works~\cite{huang2019digital,xie2018work,mark2016email,DasSwain2019FitRoutine,saha2019libra,shami2015inferring,dasswain2023focused,shami2014understanding}.}
    \label{fig:storyboards}
\end{figure}

\subsubsection{Storyboard-driven semi-structured interview protocol} 
For each interview, we first gave participants an overview of the study goals and structure. We then sent them a copy of a slide deck that included images of the four storyboards. We conducted ~60-minute video interviews over Microsoft Teams which included a screen-sharing session with the same slide deck and was audio/video recorded with the participant's consent.

Participants were informed that the study's goal was to understand how their unique position within their company shapes their attitudes toward these technologies. For each storyboard, we asked if the storyboard was relatable to them and what were their immediate reactions.Then, we asked if they had any concerns about the technology and its use depicted in the storyboard. Since our goal was to build a holistic understanding of different stakeholders' concerns, if participants could not think about any concerns, we showed them a list of potential harms from the storyboard. \edit{The list of harms was generated based on our knowledge of prior academic literature (e.g., \cite{introna2000workplace,cottrill2020sensing}), as well as brainstorming and discussion amongst the research authors.} The harms were used as a way to help scaffold thinking around potential concerns, and we clarified that the harms may or may not actually happen in practice. For participants who were Organizational governors or AI builders, we additionally asked them to ``switch hats'' to a worker role during parts of the interview protocol to uncover a broader range of potential concerns. After understanding their reasons for the concerns that they raised, we prompted participants to think about potential mitigation strategies or solutions that could address these concerns. 

For worker data subjects, we asked what they would like to change in the storyboard to help them feel less concerned. 
For AI builders, we asked them to imagine a scenario in which they were assigned the role of designing this technology, to engage them in thinking about choosing now to build it or how they would change it. 
For organizational governors, we asked them to imagine that a product team pitched the idea to them, and they were asked to provide feedback. We additionally asked about what improvements or mitigation they might recommend that cannot be addressed by a product team (e.g., policy-related mitigation). We randomized the order in which we showed the four storyboards to avoid ordering effects. After showing all four storyboards, we asked questions to help debrief on the storyboard-driven section of the protocol. This included questions about whether participants were interested in seeing the technology in their workplace, their broader reflections of the depicted storyboards, and if they had prior experiences using data-driven wellbeing technologies. We prioritized using the interview time to go through all four storyboards and only asked the post-storyboard questions if there was remaining time. We kept the overall structure and nature of questions consistent for all participants. The study protocol and materials were refined through a pilot study with three participants (1 HR and 2 builders). 

\subsubsection{Storyboard construction}
We designed the storyboards using an iterative process that involved collecting and reading research papers across UbiComp, HCI, and CSCW that propose or study wellbeing sensing technologies and sketching use case scenarios inspired by those papers. In total, we reviewed 31 research papers and selected 19 papers that represented a diverse set of wellbeing problems and methodologies (\autoref{sec:related}). We outlined and organized the use case scenarios from these 19 papers into four topical categories in a bottom-up manner: 1) Work Patterns and Scheduling (\SBWPS{})~\cite{huang2019digital,xie2018work}, 2) Notifications and Focus Work (\SBNFT{})~\cite{mark2016email,dasswain2023focused,saha2023focus}, 3) Emotions and Social Interactions (\SBSIE{})~\cite{shami2015inferring,shami2014understanding,de2013understanding}, and 4) Productivity and Job Fit (\SBJFP{})~\cite{saha2019libra,DasSwain2019FitRoutine}.  
~\autoref{fig:storyboards} shows images of all four storyboards. 
Each storyboard follows the general pattern of having four panels---the first panel shows the wellbeing problem that the technology attempts to address, the second panel shows the data collection and sensing, the third panel shows the intervention the sensing technology provides, and the fourth panel shows a positive, real-world downstream course of action that resulted from that sensing technology. For participants who were data subjects, we changed the name of the worker in the storyboard to their actual name prior to the interview. 
The scenarios depicted in the storyboards all led to a positive outcome for the storyboard character because the stories are inspired by prior literature innovating on the technology.
Our storyboards were also intentionally designed to omit certain details (e.g., the accuracy of sensing overtime work), and these ambiguities gave participants an opportunity to question or respond to these gaps.

\subsection{Analysis} 
Drawing on practices in reflexive thematic analyses (e.g., in~\cite{clarke2015thematic,braun2006using}), we qualitatively analyzed over 33 hours of transcribed interview recordings. 
All the interview transcripts were first open-coded by at least one of the researchers, while referring to the research questions and protocol topics. For most interviews, we ensured that the coder was the researcher who conducted the interview. Otherwise, the researcher who conducted the interview looked over the coding to reconcile any misconceptions. During the process, the research team met frequently to discuss the codebook and resolved disagreements such as the granularity of codes. 
After the open-coding step, we iteratively refined and grouped all the codes into successively higher-level themes. In the end, this resulted in four levels of themes. The highest level of themes is organized into the three subsections of findings presented in the~\autoref{sec:results}.

Our initial coding and grouping were agnostic of participants' roles and specific storyboards to avoid interpretation biases, while the codes were tagged with such information. Using these tags, after the first round of analysis, we looked for differences across the following two dimensions: 1) Stakeholder group and 2) Storyboard topic. We adopted inductive thematic analysis with the researchers in the team to identify themes of different stakeholders' needs, concerns, and desires with respect to these technologies. 
We noted which themes and sub-themes were unique to specific stakeholder groups and included that in the findings whenever applicable. We did not find any notable difference across storyboards, so the sub-themes presented in the Results (\autoref{sec:results}) pertain to all four storyboards. 

\subsection{Privacy, Ethics, and Positionality} 
Our study concerns the desirability and functionality of technologies that would be situated in workplaces by soliciting perspectives from participants employed at those workplaces. Therefore, our participants may not be comfortable fully expressing concerns or critiques during the study. Throughout the recruitment and the interviews, we assured the participants that their participation was completely voluntary, all questions were optional, their responses would be anonymous, and they may ask us to remove any recorded content from the study data after the interview session (no participants asked us to remove recorded content). To avoid identifiability and traceability of our participants, we analyzed the data in a de-identified fashion, and in this paper, we paraphrase their individual quotes and present demographics at an aggregated level, without disclosing organizations and specific roles of individuals. Our research team comprises researchers holding diverse gender, racial, and cultural backgrounds, including people of color and immigrants, and hold interdisciplinary research expertise in the areas of HCI, UbiComp, CSCW, and critical computing. This study was approved by the Institutional Review Board (IRB) at our research institution.

\section{Results}\label{sec:results}
\begin{figure}
\centering
   \includegraphics[width=0.85\columnwidth]{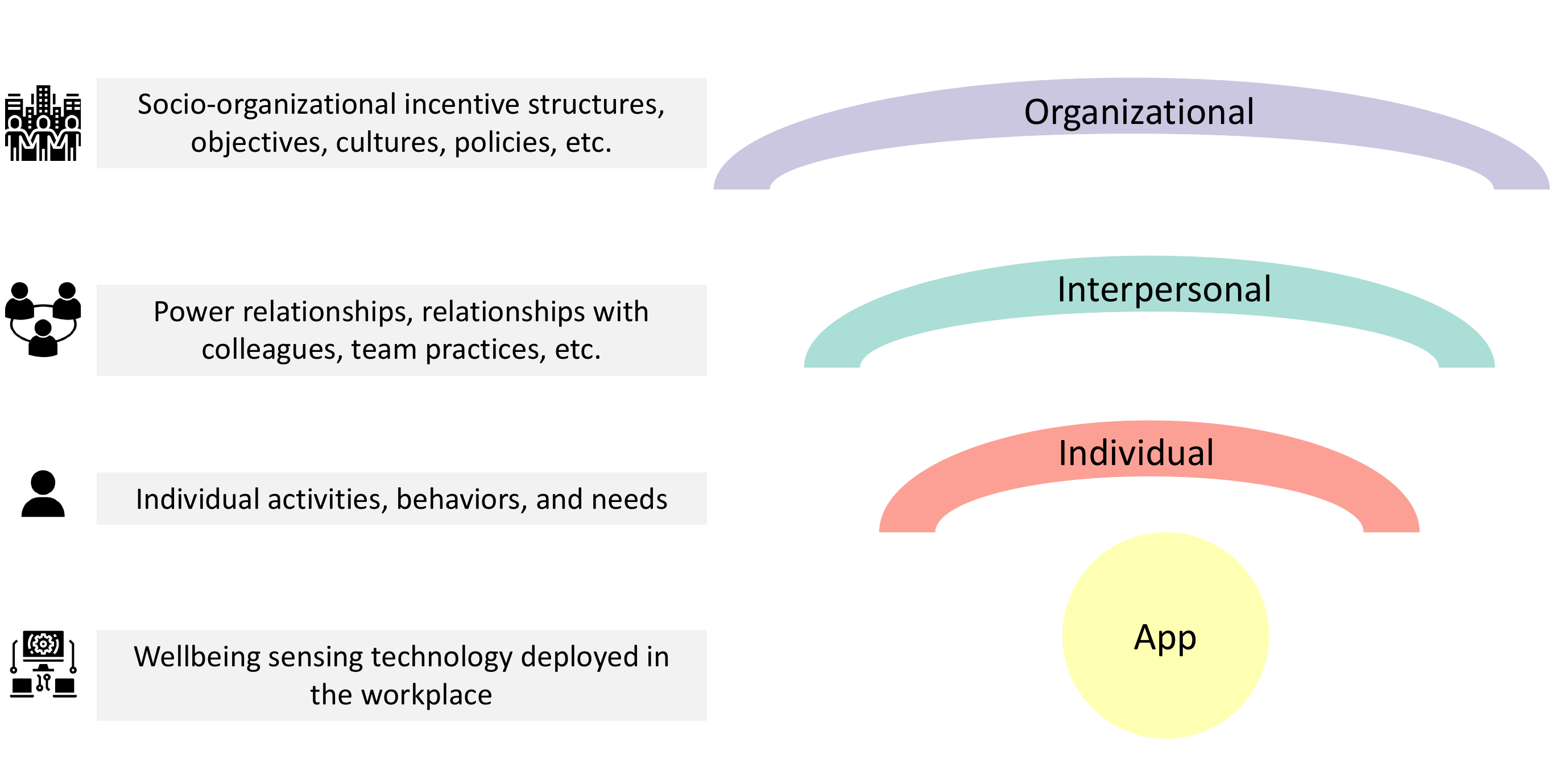}
    \caption{Workplace-specific layers overviewing where a wellbeing sensing technology is situated.}
    \label{fig:summaryFindings}
\end{figure}

\begin{figure}
\centering
    \includegraphics[width=\columnwidth]{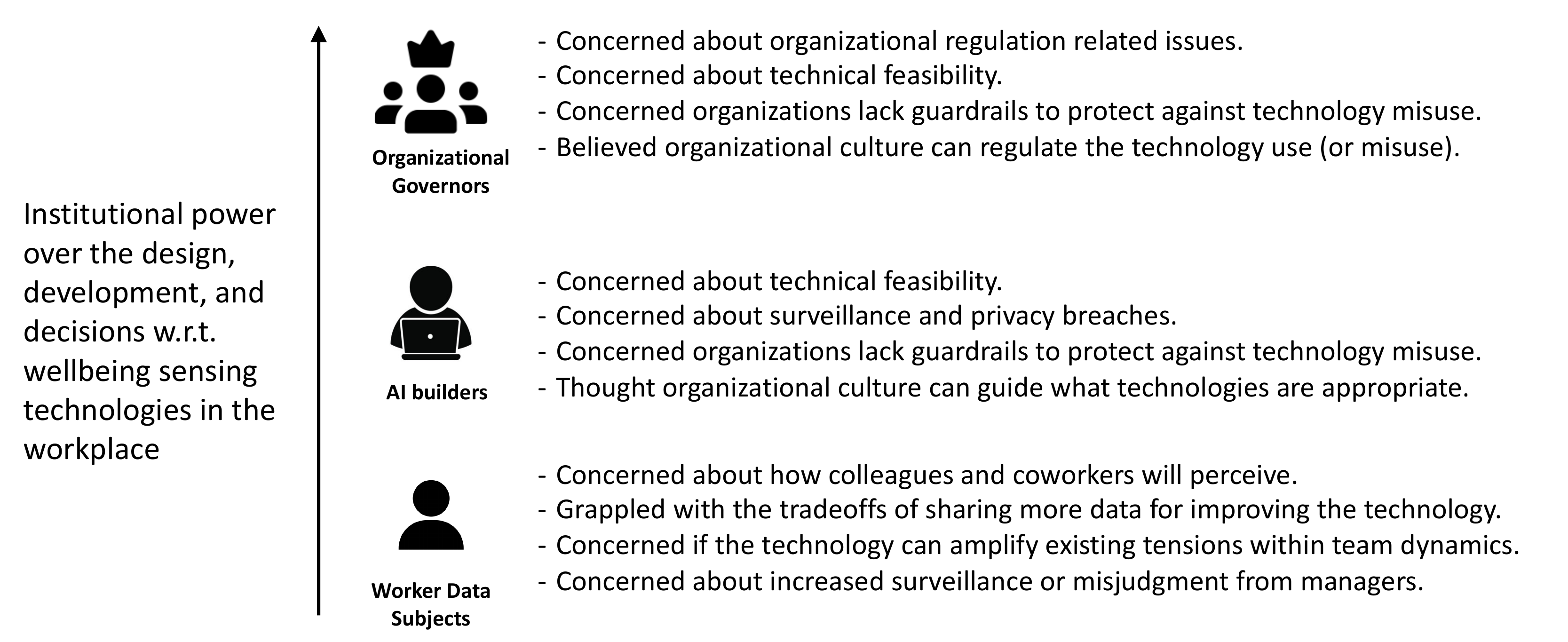}
    \caption{A figure showing example viewpoints per stakeholder group.}
    \label{fig:summaryFindingsStakeholders}
\end{figure}

Our findings surface the harms and benefits that organizational governors, developers, and end-user workers envisioned from the potential uses of data-driven workplace wellbeing technologies. We organize our findings into three sections, split into three layers cross-cutting the workplace context (\autoref{fig:summaryFindings}). At a high level, we first describe how participants envisioned how these technologies may be mediated by their existing organizational culture and workplace incentives (\autoref{sec:results:org}). Next, we discuss participants' perceptions around how their relationships with their managers and colleagues may be enhanced or harmed, depending on their interpersonal relationships and team culture (\autoref{sec:results:inter}). Finally, we discuss how participants perceived they, as individuals, may feel supported by or at conflict with wellbeing technologies deployed in their workplace (\autoref{sec:results:ind}). We note that these layers are all intertwined (e.g., concerns at the organization-level cascade down to impact individual-level behaviors and needs, while individual and interpersonal interactions with wellbeing technologies also shape broader organizational uses and culture). \autoref{table:harms_taxonomy} includes an overview of our findings and \autoref{fig:summaryFindingsStakeholders} includes 
some example viewpoints of different stakeholders

\begin{table}[t]
\sffamily
\caption{Summary of envisioned beneficial impacts (resulting benefits as imagined by the participants) and envisioned harmful impacts (resulting harms as imagined by the participants).}
\label{table:harms_taxonomy}
  \centering
    \footnotesize
    \setlength{\tabcolsep}{2pt}
    \begin{tabular}{p{0.45\columnwidth}p{0.53\columnwidth}}
    \textbf{Envisioned Benefits} & \textbf{Envisioned Harms} \\ 
 \toprule
\rowcollight \multicolumn{2}{c}{\textbf{Organizational Impacts}}\\
\tabitem Understand root causes of poor worker wellbeing\newline \tabitem Inform organization-wide practices and policies that better support workers' wellbeing\newline \tabitem Equip managers with evidence to advocate for more wellbeing resources\newline \tabitem Monitor managerial practices and potential biases & \tabitem Misuse to increase corporate benefit (e.g., increase profit and public image)\newline \tabitem Justify and over-rely on technology ``band-aids'' over systemic solutions \newline\tabitem Motivate creation of harmfully restrictive regulations \newline\tabitem Support biased inferences about reasons for (poor) wellbeing\newline \tabitem Make employment decisions unfairly target workers whose working patterns deviate from the norm \newline \tabitem Create a culture that constrains open workplace communication\\
\rowcollight \multicolumn{2}{c}{\textbf{Interpersonal Impacts}}\\
\tabitem Support proactive discussions between manager and worker about strategies to improve wellbeing \newline \tabitem Lower hesitations towards seeking wellbeing support from manager & \tabitem Enable unfair treatment from managers with conflicting responsibilities to their organization and their workers \newline\tabitem Promote evaluation of worker wellbeing as a proxy for productivity \newline\tabitem Set team practices that misalign with individual workers' wellbeing desires or needs \newline\tabitem Collect data that propagates unhealthy workplace competition and comparison \newline\tabitem Risk data leakage that harms workers' professional image \newline\tabitem Misunderstand worker activity by overlooking circumstantial or contextual factors\\
\rowcollight \multicolumn{2}{c}{\textbf{Individual Impacts}}\\
\tabitem Help identify and prevent work patterns that detriment individual wellbeing \newline\tabitem Support workers in enforcing individual work practices based on personalized notions of wellbeing & \tabitem Create unhelpful or offensive recommendations that fit an over-simplified notion of workplace wellbeing \newline\tabitem Overlook workplace incentives or over-attribute problem sources to the workplace\newline\tabitem Gaslight or patronize from suggesting a sensed wellbeing state that misaligns with workers' perceived wellbeing\\
\bottomrule
\end{tabular}

\end{table}

\subsection{Organization-Level Impacts}\label{sec:results:org}
\subsubsection{Enabling Improved Company-Wide Wellbeing Practices and Policies through Wellbeing Data} \label{sec:result:org_pos}
Some participants envisioned that the data from wellbeing technologies, even when intended for individual employees, could additionally be used by leadership to understand how the organization can support its employees' wellbeing through other resources. For example, some organizational governors described how anonymized wellbeing data aggregated across all employees could inform the design of organizational policies that target structural challenges impacting their employees' wellness. As one organizational governor (G11) put it, ``[Human Resources] can take all of this information from all of the people in the company and see if there's a pathology in the organization that can be fixed because of some patterns they're seeing across multiple peoples, and that is going to be applied to improve the life of everybody.'' 

Other participants envisioned that wellbeing apps could support managers in communicating to leadership any need for additional resources to support their employees' wellbeing. For example, one participant described that collecting data on wellbeing proxies in the workplace (e.g., amount of overtime work) to identify and mitigate sources of team-level challenges: ``if everyone's doing a lot of overtime work, that probably means that you might need to ask for another spot on your team to allow more collaboration or [understand if] it's saying, `why is everyone working so hard?''' (W07). Another participant (A7) envisioned that the organization could also use the wellbeing app's data to monitor managerial practices, for example, to ensure that individual employees are treated fairly by their managers. This participant described how organizational leaders could track whether managers were distributing employee rewards that were commensurate with different workers' productivity, as inferred by data collected in the wellbeing app.

\subsubsection{Concerns of Organizational Incentives for Misusing Wellbeing Technologies} \label{sec:result:org_neg1}
Some participants expressed skepticism towards the wellbeing sensing scenarios described in the storyboards, believing that the implementation would benefit the employer more than the individual worker themselves. Participants felt that organizations are incentivized to ``maximize profit [..] to get more out of workers'' (A7, G6), making it likely that wellbeing apps—even when deployed with a stated intention to improve individual wellbeing—may be another workplace tool to increase the organizations' profits. 

Participants also perceived irony in having profit-maximizing organizations collect their employees' data and claim ``for employees' welfare'' (A12). One organizational governor (G7) perceived the wellbeing apps as ``palliative care to make people feel better about the situation, [..] putting band-aids instead of looking at the systemic issues that might be the root cause of many of these.''
Other participants expressed skepticism that organizations may collect employees' wellbeing data to boost their public image (A5) by providing data that allows them to publicly claim their employees were benefiting from positive mental health (A9). 

Participants envisioned a range of ways the organization may re-purpose employees' wellbeing data for reasons beyond supporting their wellbeing, like enabling new forms of corporate regulation. For example, one AI builder was concerned that organizations might use data from employees' working hours (Storyboard \SBWPS{}) to identify those who are using their work devices for personal reasons, risking individual workers' employment. 

Other participants noted the sensitive nature of wellbeing data, making it susceptible to misinterpretation that is harmfully biased towards different groups of workers, even when seen in aggregate. For example, one organizational governor cautioned that wellbeing data could be used to make biased causal inferences: 

\begin{quote}  
\small
\textqt{``women might feel particularly dispassionate or unproductive. Let's say, if they're experiencing their periods [..] And so then you'll see these patterns where ohh, women are less productive than men and women are less passionate about their jobs than men. Yeah, because women have a few days a month where, you know, they're feeling like dying, but they're working anyway.''} (G7, Storyboard \SBJFP{}). 
\end{quote}   

Further, participants envisioned that workers might begin altering their workplace behavior to protect themselves against unwanted data collection. For example, one participant described that workers may feel an increased ``level of consciousness [.. because] they know their messages are being analyzed, or you know, peeped in.'' (W4). Another participant similarly described how undesirable data sharing, even for the intended purpose of improving one's wellbeing, could ``constrain someone's communication pathways'' (G4), introducing new barriers to supporting open communication with managers. This aligns with the paradoxes of behavioral visibility in organizations as identified in~\citet{leonardi2020behavioral}, \edit{particularly the connectivity paradox, i.e., the ubiquity and wide availability of workers' digital traces, efforts can lead to a work environment of overcommunication, interruptions, and interference~\cite{leonardi2020behavioral}}.

\subsubsection{Reliance on Organizational Culture as a Regulatory Force}\label{sec:result:org_neg2}
Without effective regulations in place, organizational governors and AI builders described how the organization's culture---for example, its ``ethics and code of values'' (A10, Storyboard \SBNFT{})---could serve as a compass to signal what forms of wellbeing apps may be appropriate. For example, one organizational governor described that deciding whether to deploy a given wellbeing technology should consider a ``creepiness factor''---where creepiness arises when ``we cross the threshold from being someone's employer to being more than that'' (G4). 

Importantly, organizational governors noted that companies might be more likely to rely on culture alignment-based assessments (e.g., like the ``creepiness factor'') when there are no existing regulations. For example, when discussing the factors that determine whether a given wellbeing technology could get deployed or not, one organizational governor described how considerations might differ across countries. Comparing the U.S. to Europe, where there is stricter legislation, the participant described: ``In Europe, this should probably be a no-go. In the U.S., I think for some employers, it would probably depend on the corporate culture and tone that they want to set.'' 

While some organizational governors described assessing for ``creepiness'' (e.g., as in~\cite{ghoshray2013employer,watkins2007workplace,seberger2022still}) to understand what forms of data-driven wellbeing technology may be appropriate given the existing culture within their organization, worker data subjects similarly centered their own understanding of their organization's values to inform their judgments of wellbeing technology: ``I am not a supporter of this technology [...] My company has a great value of privacy, and this would be a complete breach of its values'' (W4). 

However, other participants described the risks of relying solely on organizational culture as a compass for understanding what forms of technology may or may not be appropriate to deploy. For example, organizational governors described the fluid and often volatile nature of organizational culture, given workplace norms can shift quickly with leadership changes. Another organizational governor (G5) described that existing norms in the workplace might not be favorable towards workers, with practices differing based on how ``traditional'' a company is: 

\begin{quote}
\textqt{``The more modern enterprises, I would keep [Human Resources] out [of employees' wellbeing data]. The more traditional enterprises are more structured. There's a processed way of doing everything: [… Human Resources] is instrumenting everything. They are pretty much in the loop and [in] the know-how on every career decision that is being made about the employee, employer, and manager''} (G5).
\end{quote}

Another AI builder (A7) also expressed concerns that, without appropriate regulations, harms to the workplace culture and workers themselves may be less visible: ``There are power structures in the company that could lead to this being used to the detriment of the employees, fairly easily without anyone even noticing.'' As an example, they told us how individual employees currently have no effective ways to raise concerns about their managers to higher-ups, creating risks of managers misusing these apps without others knowing. While participants generally acknowledged the importance of having company-wide policies to govern the use of wellbeing technologies, they also warned that company-level policies may still unfairly disadvantage workers. For example, one organizational governor (G7) admitted, ``It's very difficult to have an employee-employer policy that is fair to both parties. [It's] most likely to be unfair to the employee and fairer to the employer, because they have the power they have – they'll be the ones or making the policies for the company itself.'' Instead, one organizational governor with a legal background voiced a need for new legislative policies, specifically for the worker wellbeing domain: 

\begin{quote}
\textqt{``[There are] differences between a labor question or a data protection question [..] should definitely be combined into something that is for employee wellbeing and employee data. [Labor laws and data protection laws] are two different areas of law and policy. [..] And it's challenging to enforce it to have some rights for the worker.''} (G6)
\end{quote}

\subsection{Interpersonal-Level Impacts}\label{sec:results:inter}
\subsubsection{Enhancing Manager-Worker Relations} \label{sec:result:inter_pos}
Many worker data subjects expressed excitement about using the wellbeing sensing app to enhance their communications with their manager because they resonated with the challenges faced by the character in the storyboard. A participant approached the storyboard with light-hearted curiosity or humor, saying that they would find it ``cute and hilarious'' if their manager discussed their work patterns with them based on the wellbeing app (W7). Looking past the light-heartedness, participants imagined ways the portrayed sensing app could improve day-to-day work, especially in equipping managers to support them in being more reflective and proactive about their wellbeing.

For example, participants envisioned that having an app share their work hours with their manager could open opportunities to discuss strategies to improve their wellbeing. In some cases, participants imagined that the wellbeing app could help them share concerns with their manager around sensitive topics that are currently challenging to express by themselves. For example, one participant described that their cultural background imbibes shame in bringing up wellbeing-related challenges with their manager. This participant envisioned that a wellbeing app that documents and shares working hours could help nudge her towards having this conversation: ``I think half the battle is bringing up the issue. This app that actually has the data and brings it up to your manager, prompting you to talk to you [..] So it's helpful in that sense.'' (W10, storyboard WPS).

\subsubsection{Concerns Around Using Wellbeing Data for Team Management} \label{sec:result:org_pos}
Participants in our study imagined ways workplace actors of the same position could leverage the same wellbeing app to either harm or help workers' wellbeing. Participants particularly viewed managers as holding a dual responsibility: bearing services towards their organization to evaluate their reportees' performance and---in tension with it--service towards their reportees to support their wellbeing. Moreover, with the workplace context ambiguating what it means to have ``positive wellbeing'' versus ``good productivity,'' workers were concerned that efforts towards improving their wellbeing could easily be misguided into evaluating their wellbeing as a proxy for productivity. Because of this, they feared that wellbeing data would lead to undesirable or unfair treatment from their managers. 

Worker data subjects described how even well-intended managers might unknowingly use the wellbeing app against employees' best interests. As one worker data subject (W10) described: ``it doesn't matter how great your relationship with your manager is, it's very difficult to overcome your own perception of what is what constitutes a good worker.'' Another worker data subject also described that managers and employees may have misaligned notions of what preferable levels of ``work-life balance'' should look like: 

\begin{quote}
\textqt{``it's good that the manager has this information, but if there was, say, a new project, I really wanted to work on. The manager might be like, `Oh, well you've got too much on your plate, and I don't wanna give it to you.' '' (W8)}
\end{quote}

Other participants were concerned that tracking work hours, even for wellbeing-related purposes, could trickle into setting potentially harmful blanket standards for working hours amongst their team members. For example, worker data subject (W3) described how the wellbeing app might amplify existing tensions amongst colleagues who may feel pitted against each other at their workplace: ``[collecting working hours] may not pan well, in places where there's pressure to work more hours and compare against other employees'' (W3, Storyboard WP). Moreover, participants expressed concerns that wellbeing apps used to manage their team could pose risks to setting boundaries with their colleagues. Given that the wellbeing app may sense more personal attributes related to their emotional wellness (e.g., in Storyboard \SBSIE{}), one worker described that any data sharing or data leakage could undermine the professional image workers may want to uphold in front of their colleagues: ``[I would be] very concerned if this gets shared with others since in the workplace [..] if [a worker] wants to convey a certain image of how she feels about her work. This could impact whether she gets promoted'' (W10). 

Participants additionally felt concerned that managers might misunderstand workers' wellbeing or productivity, by overlooking critical factors not captured by the wellbeing app. For example, one participant described that a wellbeing app that uses the number of hours worked by an employee would shift the manager's attention towards the quantity of work hours, rather than the quality of the work or circumstances influencing the working hours. As one worker described: 
\begin{quote}
\textqt{``when a manager doesn't get to see their folks directly, and now has to differentiate three employees who are all great, but then thinks: `Because I have this [wellbeing app] and it tells me [this worker] is working lot harder – longer hours than if he needs to, so he deserves a better bonus.' [The manager could make this decision] without taking into consideration, `why is he working longer hours? Is he being productive those hours or is he just connected and chatting with his friends on Facebook?'''} (W6)
\end{quote} 

Participants stressed that the wellbeing app should collect a worker's unique circumstances and richer contextual data. 
For example, a participant mentioned the importance of supplementing any data with more context: 
 \begin{quote} 
\textqt{``If somebody's not engaging in a particular task, it could be for a number of reasons, right? [..] There [could be] some personality personnel issues or it could be, something in their life that's intruding on that particular thing. Or maybe there's something [with] different time zones. [..] It's just you never see any of that context with this technology. So I think context needs to be provided somewhere.'' (W2, Storyboard \SBJFP{})}
\end{quote}

To collect contextual data, participants suggested data collection through active means, not just passive sensing, could serve as a complementary source of information. For example, one data subject (W2) was open to inputting their wellbeing data manually in conjunction with the passive data collection: ``It could ask what is your current feeling right now, it could use [that with passive sensing] to say maybe something is up. I think that kind of makes good sense'' (W2). 

Overall, participants from all three stakeholder groups also stressed the importance of having policies that required managers to undergo training processes to support them in appropriately using the wellbeing app. For example, participants described that managers should learn what data is and isn't captured by the sensing inferences, so they could more accurately interpret the data insights before making decisions aided by them.

\subsection{Individual-Level Impacts}\label{sec:results:ind}

\subsubsection{Enabling Positive Workplace Behaviors} \label{sec:result:ind_pos}
Some participants described that the wellbeing app could help identify and prevent working patterns that might detriment their wellbeing. For example, reflecting on their earlier months of employment, one worker envisioned having a similar wellbeing app could have helped them mitigate over-work and burnout: 

\begin{quote}
\textqt{``I was a computer engineer by degree, but I joined management, so I had to prove to people that I could do it, as a lot of questions were being raised by people around me. So I took up projects that weren't normal. I took on so many projects that I gradually started burning out to the extent that I was ill with typhoid, but was still working. I got some escalations through senior leadership, and then I had to deal with them [..] I think if technology could have helped me [..], wouldn't have been in that worst place in the 1st place''} (W04, Storyboard WPS).
\end{quote}

To support such positive behaviors, one participant (A11) advocated for designing technologies that empower individual workers to enforce their desired workplace norms: ``It's like humans are adapting to the norms of the systems versus the other way around. And so, how do we help the users build those boundaries? And the technology adapts to that versus the other way around.''

\subsubsection{Concerns about the Prescriptive Nature of Recommendations} \label{sec:result:ind_neg1}
Participants were concerned that their individual needs could be overlooked by the recommendations provided by the wellbeing app. They pointed out that wellbeing recommendations may fit an overly simplified notion of wellbeing in a workplace context, causing discomfort to the worker who receives the recommendation. In particular, they described how their ``workplace wellbeing'' may be impacted by personal challenges unrelated to their work and workplace-specific incentives. Yet, wellbeing recommendations may not disambiguate these factors, causing discomfort or other harms to the individual worker. 

Some workers were concerned that wellbeing-related recommendations might provide suggestions that overlook, or cause tensions with, career-related incentives that may not help with their work performance. For example, A6 described that a recommendation to take a break might be harmful to an employee if their goal is to get promoted; however, it could be valuable if they want to receive help forming a more sustainable working pattern. Without understanding the worker's preference in a given moment, the recommendation in the wellbeing app appeared to assume a preferred tradeoff between their goals for wellbeing versus career progression. 

Moreover, participants were worried that recommendations might be superficial or ineffective because the wellbeing app cannot delineate whether their problems are caused by workplace-specific versus personal factors. A10 was concerned that in Storyboard \SBWPS{}, the app might recommend an individual to go back home after work hours when they purposely work overtime as they do not have a happy domestic life at home, exacerbating their domestic life further. This participant contextualized this case with women in India, ``women are 35\% of our workforce, [..] one of the primary reasons women work in the workplace is to avoid their family issues at home.'' 
Relatedly, workers described feelings of discomfort with a wellbeing app that suggests workplace-specific tips for improvement because sensed patterns of lower wellbeing may also be signs of broader or more personal challenges the worker is facing: 

\begin{quote}

\textqt{``So you're jumping to a solution without knowing the root cause. I might be tired because I'm physically sick, or because my work bores me, or because I have a small baby that's teething, right? It's just could be so many things so. Energizing tips are unlikely to meet the root cause of the problem.''} (W6, Storyboard \SBNFT{}).

\end{quote} 

\subsubsection{Harm from Misalignment between Workers' Perceived Wellbeing and Sensed Wellbeing} \label{sec:result:ind_neg2}
Participants across all the stakeholder groups expressed concerns about seeing wellbeing recommendations that infer an emotional state that misaligns with their own understanding of their wellbeing. For example, participants imagined workers may feel emotional distress from receiving a recommendation similar to that described in Storyboard \SBJFP{}. Because the wellbeing recommendation pertained to employment-related concerns, W8 described this could cause ``panic.'' An AI builder similarly described how it might lead to feeling ``gaslighted'' by the app:  

\begin{quote}  
\textqt{``[..] If I'm doing well in my job, I would not want an AI to tell me: `Hey, based on your number of completed tasks, [..] you may be struggling with some aspects of your work.' Excuse me, I know I'm doing a good job, and it feels like I'm being gaslit by this technology.''} (A2, Storyboard \SBJFP{}).  
\end{quote}  
Similarly, another participant described that receiving wellbeing recommendations in a workplace context risks feeling ``patronizing'' (A5, Storyboard \SBSIE{}), given they or other workers may feel capable of identifying their own needs for wellbeing. To mitigate these concerns around providing inappropriate wellbeing recommendations, participants in all three stakeholder groups suggested that the wellbeing app instead connect them to human-based support so that the worker may receive help more collaboratively improving their wellbeing, rather than attempting to give a recommendation itself.

\section{Discussion}

Amidst multiple transitions in the workplace, including remote and hybrid work, organizations are increasingly implementing data-driven technologies that track workers' behaviors and actions. In parallel, there is growing corporate interest in adopting technologies that aim to improve workers' wellbeing. A recent survey estimates worker wellbeing in the US costs to be $\sim$20.4 Billion USD in 2021, which is forecasted to grow four-fold in the next five years~\cite{forbes2021Wellbeing}. Existing research and development on wellbeing technologies that use passive sensing have often focused on interactions between the individual user and the technology. However, growing corporate interest in adopting sensing-based wellbeing technologies for the workplace raises new questions about how such technologies may interface with complex organizational structures, interpersonal dynamics, and individual preferences. While there is an emerging line of work shedding light on data subjects' concerns and desires towards various forms of wellbeing-related workplace technologies (e.g., on Emotion AI~\cite{corvite2022data}), we lack an in-depth understanding of how various organizational stakeholders envision these technologies may harm or benefit their workplace and selves. 

In this study, we form a systematic understanding of the beneficial and harmful impacts that stakeholders envision at the organizational, interpersonal, and individual levels. To do so, we include participants across multiple roles---from organizational governors who benefit from a higher-level perspective on the governance needs and organizational goals of such technologies to worker data subjects who are often most directly impacted by the deployment of such technologies (yet have the least say in influencing the introduction of emergent technologies). Overall, our findings surface multiple concerns targeting ambiguities and misalignments around 1) the core definition of workplace wellbeing, 2) underlying technical limitations in how well sensing-based technologies can scale to meet individual workplace wellbeing preferences, and 3) inadequate policies and reliance on organizational culture for governing the appropriate use of such technologies.

\subsection{\edit{Contesting Notions of ``Workplace Wellbeing''}}
\edit{Our findings suggest that designing truly worker-centered workplace wellbeing technologies that are governed by organizations may be a \textit{wicked problem}~\cite{roberts2000wicked,park2022designing}---a goal that is difficult or impossible to solve.} Across all three layers of organizational, interpersonal, and individual impacts, this study provides empirical evidence that what constitutes a ``workplace wellbeing problem'' is not easily definable within an organization, between workers with interpersonal relations, and across individual workers. In our study, participants had varying perspectives on whether certain behaviors and goals (e.g., feeling higher levels of satisfaction while working) were related to what they would define as ``workplace wellbeing.'' Moreover, these perspectives may not necessarily be moderated by their role in the workplace; they may also be manifested by individual differences in underlying traits, life experiences, and perceptions. \edit{For example, participants described ways in which their gender, household events, prior experiences with managers, country of residence, and ``modern''-ness of their company shaped their perceptions around what goals and practices around ``workplace wellbeing'' resonate with them. In one case,} 
\edit{a participant described that, because of the cultural norms she grew up with, she is not inclined to have ``workplace wellbeing'' conversations with her manager. In this case, the participant appreciated that this norm was embedded into the hypothetical technology presented in the storyboard, as she imagined it could nudge her towards having more honest communication around her wellbeing needs with her manager (\autoref{sec:result:inter_pos}). However, as other participants described, practices around having wellbeing-related conversations with employers may not be universally desirable, for example, in settings where there is less certainty that this information will not shape evaluations of worker performance. }

\edit{Interestingly, this example also introduces complexities to interpreting differences between workers' individual goals for ``workplace wellbeing'' and the goals purported by the technology. 
On one hand, differences can be presented as an opportunity to nudge more proactive wellbeing practices; however, these same differences can also frame technology as a mechanism to control and manipulate worker behavior (as other participants described in Section \ref{sec:result:org_neg1} and \ref{sec:result:ind_neg1}). Given challenges to actually ensuring collective value alignment in practice, and that (mis)alignments cannot be easily categorized as ``good'' or ``bad'', future work should center the \textit{context- and worker-specific implications} that arise from potential differences between workers' desired notions of ``workplace wellbeing'' and constructs used by technology.} 

\subsection{\edit{Tensions to Sensing and Scaling ``Wellbeing'' in the Workplace}}
\edit{Relatedly, participants' responses reflected that their perceptions of what constitutes appropriate definitions of ``workplace wellbeing'' cannot be divorced from the underlying \textit{intention to build technology} around workplace wellbeing.}
Indeed, prior work by \citet{docherty2022re} has discussed how, in the context of digital wellbeing measurement, notions of wellbeing are not easily generalizable and scalable. \edit{Our study supports and extends these arguments, suggesting how workers' perceptions of what should constitute a ``workplace wellbeing problem'' are often shaped by the implications tied to \textit{measuring} a given ``problem.'' For example, participants were wary of notions of ``workplace wellbeing'' that involved their organization collecting data on their working hours, given the lack of effective guardrails for mitigating risks of corporate misuse and data misinterpretation (\autoref{sec:result:org_neg1}). Measuring ``workplace wellbeing'' also introduces new tensions when scaled at different levels---from individuals to organizations.} For example, we found additional harms may arise from assuming certain notions or factors contributing to ``workplace wellbeing'' that align with organizational goals but not individual worker or team needs. Defining organizational wellbeing targets that do not align with workers' wellbeing goals may produce misleading outcomes and decisions on how to better support workers. Yet, many existing workplace wellbeing technologies justify their reliance on passive sensing and algorithmic inference approaches due to their perceived ``objectiveness'' and scalability, along with convenience and unobtrusiveness~\cite{dasswain2020social,mirjafari2019differentiating}. \edit{Relatedly, workplace wellbeing sensing technologies may narrowly focus on a specific wellbeing problem (e.g., sedentary behavior at work, working overtime). Our findings strongly suggest a need to understand the effectiveness of sensing wellbeing as subjective upon an individual's context-specific needs and desires for supporting their wellbeing. Future work should explore the design of new interventions that mitigate tensions between organizational incentives for scalable technologies and worker needs for personalized specific proxy measures for wellbeing that meet their individual needs.} 
 
Relatedly, participants in our study also surfaced the importance of understanding the \textit{context} surrounding their workplace behaviors and wellbeing needs. This need was especially pertinent when considering how their managers may use this information to inform team decisions. Moreover, rather than relying solely on passive sensing technologies for identifying patterns in workers' wellbeing, participants expressed a desire to explore how \text{active} participation can support the design of improved wellbeing technologies. On the one hand, there is an opportunity to explore how workers may be empowered through self-reported, qualitative wellbeing documentation. These methods can simultaneously be unobtrusive and relevant while providing workers with additional agency and control over algorithmic inferences. At the same time, however, such technologies should be governed with considerations around how they may increase risks of privacy and anonymity breaches in the workplace. Moreover, prior literature on the privacy paradox~\cite{barnes2006privacy} and empowering resignation~\cite{seberger2021empowering} suggest that enabling greater customizability and control over data collection may in fact lead to increased resignation over data sharing decisions. Future research should explore whether and how personalized wellbeing technologies, combining active and passive mechanisms, could, if at all, help mitigate some worker burdens without exacerbating others. 

\subsection{\edit{Complementing Technology Design with Organizational Policy Design}}
Across stakeholder groups, but especially amongst organizational governors and builders, participants raised concerns around the lack of–and the critical need for–regulations and policies that help govern the responsible implementation and use of workplace wellbeing technologies. Participants envisioned ways and reasons for which wellbeing technologies designed to meet the individual needs and desires of workers could still be irresponsibly used by those in higher positions of power (e.g., leaders, and managers), after deployment. For example, participants envisioned that existing organizational policies could further disadvantage workers and prioritize organizational needs (e.g., ensuring workers cannot begin lawsuits against the company). Participants also described that workers subject to these technologies may not always have the leverage within their organizations to meaningfully communicate and report when they believe their data is being used unfairly. \edit{Moreover, as discussed earlier, regulations have the potential to help mitigate some risks but, at the moment, heavily depends on the country (e.g., governance across European and US-based organizations differ).}

\edit{These findings suggest that ensuring wellbeing technologies truly center worker needs in practice necessitates considerations beyond understanding workers' needs and concerns towards the technology. HCI research on workplace technologies commonly focuses on improving technology use and design (e.g., understanding workers' use of technologies or designing new workplace technologies). With the rapid rate at which workplace technologies are being innovated and deployed, there is also an opportunity and need for new approaches that enable workers to proactively inform the design of \textit{organizational policies} that govern technology use. A growing body of work across human-computer interaction and policy has suggested that considerations around technology design and use are deeply intertwined and should be accounted for alongside policy considerations~\cite{jackson2014policy, wong2015wireless,centivany2016popcorn}. Yet, in current practice, policy considerations are often an afterthought, holding secondary priority to the design of new technologies. As a result, emerging technologies are often deployed without sufficient policies to proactively prevent harms; instead, new policies may be created to ``clean up'' after harms that occur \textit{after} a technology is already deployed~\cite{jackson2014policy}.}

\edit{Our study suggests that organizational stakeholders, when scaffolded through storyboard-driven interview questions, are able to articulate and envision various harms that can inform the creation of new policies, in addition to improving the design of technologies. For example, participants anticipated that they may be able to more proactively mitigate harms from technologies through deliberate decisions about \textit{whether} to deploy a given technology. They anticipated risks arising from their current organization's lack adequate guidelines for informing these decisions; as a result, some participants envisioned leaders may default to informing decisions based on a ``creepiness factor'' that assess violations to workers' workplace expectations (Section~\ref{sec:result:org_neg2}). These risks that participants identified raise the need for organizational policies and guidance for more systematically informing decisions about \textit{whether} a technology is appropriate for deployment. Other participants described implications for policy to improve technology use, for example, through policy that creates pathways for workers to share feedback on their managers' use of  wellbeing data and report inappropriate uses (Section~\ref{sec:results:org}). Indeed, building on literature reframing policymaking as a design problem~\cite{junginger2013design}, our study suggests design approaches like the storyboard-driven interviews we used may provide a promising approach to more proactively foresee harms, supporting the worker-centered design of technology and policy simultaneously. Future work should continue to explore approaches for the design of organizational policies and workplace technologies that can, together, help ensure that new workplace technologies are designed and used responsibly~\cite{yang2023designing}.}

\subsection{\edit{Designing for Choice and Refusal in Workplace Technologies}}
\edit{Participants in our study often raised broader questions around \textit{whether} the portrayed technology provides an appropriate approach to addressing certain wellbeing problems. For example, one participant perceived the technology as ``palliative care [...] putting band-aids instead of looking at the systemic issues'' (Section~\ref{sec:result:org_neg1}). Other participants questioned the value of wellbeing recommendations that may overlook ``root causes'' of a worker's wellbeing challenges (Section~\ref{sec:result:ind_neg1}). Indeed, these findings suggest that technology-driven recommendations that promote improved wellbeing practices--even when the practices themselves are valued by workers--may not be effective if the actual organizational culture in which the worker and technology reside does not support workers' wellbeing. As participants feared, an organizational culture that deprioritizes workers' needs may also increase the risk that technologies designed for wellbeing are repurposed by the organization, for example, to track and evaluate worker performance. Therefore, ensuring that workers have the flexibility to use workplace wellbeing technologies to varying extents (including choosing not to use them at all, as some participants preferred) is critical to ensuring that the deployment of these technologies serve workers in practice. }

\edit{Prior work has discussed opportunities for researchers and designers to critically assess needs for supporting technology refusal~\cite{baumer2011implication,abebe2020roles,barocas2020not}. Literature on workplace tracking technologies, in particular, has furthered this discourse by exploring whether and how workplace tracking technologies can empower workers through collective data sensemaking processes ~\cite{holten2021can}. Our study further raises questions around the responsibility of researchers in supporting worker refusal of workplace wellbeing technologies; moreover, participants raised concerns about how much power workers will actually have in refusing technologies deployed by their organization. For example, policies requiring an option to opt out of technology use may not align with organizational goals. While laws and regulations may be a reliable way to ensure that organizations provide workers with an option to refuse technology, the availability of this safeguard depends heavily on the country. When the option to opt-out is unavailable, workers have historically found ways to ``work around'' technologies. For example, \citeauthor{khovanskaya2019tools} discusses ways labor unions can play a central role in repurposing workplace data collection to empower and advocate for change amongst workers~\cite{khovanskaya2019tools} and opportunities to design improved data tools for these purposes~\cite{khovanskaya2020bottom}. Future work should continue exploring these avenues for refusal and recourse for data-driven workplace technologies, even when the technology is designed and framed as a tool to help workers. Additionally, future work should explore ways to support workers in making more informed decisions around the use of data-driven workplace wellbeing technologies by scaffolding deliberations around the individual, interpersonal, and organizational benefits, costs, and tradeoffs to anticipate.}

\subsection{Design Recommendations}
Based on our findings, we provide design recommendations for both organizations interested in developing wellbeing technologies and researchers designing or studying these technologies. However, we again note that designing truly worker-centered workplace wellbeing technologies owned by organizations may be a wicked problem~\cite{roberts2000wicked,park2022designing}, in this setting, due to organizational incentive structures, power differentials, and individual values that may vary widely across workplaces and workers. Therefore, we caution that these recommendations are only appropriate when there is good reason to believe data-driven wellbeing technologies may bring more benefit than harm to workers, in practice. \edit{Moreover, reflections around \textit{whether} a given data-driven wellbeing technology is likely to bring more benefits than harm (and to whom) should be iteratively revisited, throughout the technology design, deployment, and maintenance process.} We additionally include recommendations \edit{that open opportunities to explore} how non-technical worker wellbeing support may complement or replace workplace wellbeing technologies. 
\begin{itemize}
\item Deliberate with impacted stakeholders \edit{(e.g., worker data subjects)} what notions of ``workplace wellbeing'' they feel are appropriate and desirable to target through the use of technology, before deciding to move forward with designing.  Begin by understanding impacted stakeholders' workplace-specific boundaries and needs~\cite{caglar2022user}, to form a design space for collaboratively ideating on both potential technical and non-technical solutions. Avoid design ideation driven by the availability of technical capabilities alone.  \edit{Additionally, facilitating these deliberations alongside those who typically shape decisions around the deployment of new technology (e.g., organizational governors) can help ensure that design thinking is accounts for organization-level considerations and interventions will actually be used in practice.}
\item Consider how the design of workplace wellbeing technology may shift responsibility for wellbeing challenges across individuals and the organization. Certain design decisions (e.g., about the types of wellbeing inferences or recommendations) may risk placing unwarranted blame and responsibility on individual workers, as voiced by participants in Sections \ref{sec:result:ind_neg1}, \ref{sec:result:ind_neg2}, \ref{sec:result:org_neg1}. 
\item Ensure that wellbeing technologies are not deployed as a ploy to promote ``workplace problem washing,'' by considering carefully whether the root problems of workplace wellbeing challenges may be addressable through technology, or should be complemented or replaced by other forms of non-technical, care-based human support. 
\item Provide individual workers with the power to decide ~\textit{whether, when,} and ~\textit{with whom} they would like to share any collected wellbeing data. Given that social relations with colleagues, managers, and the organization differ for each individual, workers may have different desires for (not) sharing certain data. Also, find ways to ensure that workers have an opportunity to consent and take back consent, using methods that minimize the influences of potential external incentives and pressures (e.g., placed by workers' managers or leadership, as described in Section \ref{sec:result:org_neg1}). 
\item Rather than relying solely on quantitative and passive data collection, consider how qualitative and actively collected data could provide complementary insights to workers while being less prone to causing harm. Qualitative documentation of their wellbeing and related practices may help workers better understand the context surrounding trends in their overall wellbeing, as well as aid the interpretation of data-driven wellbeing inferences. 
\item Involve impacted stakeholders, not just in the design of the wellbeing technology, but in informing the policies and regulations that govern the use of the technology. Even if wellbeing technologies are designed to support workers, once deployed in practice, they could still be used in harmful ways if governance mechanisms disadvantage workers (as described in \autoref{sec:result:org_neg2}). Ensure there are infrastructural mechanisms in place that provide workers with opportunities for recourse to potential inappropriate corporate or managerial uses of the technology (e.g., by developing worker unions). 
\end{itemize}

\subsection{Limitations and Future Directions}
Our work has limitations, which also suggest what we believe to be important future directions. Our work studies and gathers perspectives from information workers, and we cannot make generalized claims about the impact of and perceptions towards sensing technologies for other kinds of work and workplaces. \edit{Moreover, this study examines stakeholder envisioned harms, cross-cutting multiple levels of an organization, around future wellbeing sensing technologies. As wellbeing sensing technologies are introduced to real-world workplace settings, there is an opportunity for future work to more directly identify and unpack the organizational and interpersonal impacts that wellbeing sensing technologies may have, by studying their deployments in-context.} In addition, the design of our study protocol included explaining to participants the objectives of the study, including being transparent about our motives to understand their perspectives on hypothetical and envisioned technologies in the workplace. It is possible that some participants may have had an unconscious bias about implicitly interpreting that the study was about a ``product-testing.'' This may have led to responses that were over-optimistic about the use of the depicted technologies. Furthermore, the ordering in which the storyboards were shown may have played a role in influencing participants' perceptions (e.g., participants sometimes gave relative responses such as ``This storyboard is better than the previous one''). While we shuffled the order of storyboards shown to each participant to help mitigate this risk, that does not eliminate potential problems with relative perception bias.

\section{Conclusion} \label{Conclusion}
Workplace settings surface new complexities that may complicate the in-practice desirability and responsible use of these technologies. Moreover, these technologies implicate a broad range of stakeholders---including those who are impacted by and impact the deployment of, these technologies. In our study, we conducted storyboard-driven interviews with 33 stakeholders across three participant groups: organizational governors, AI builders, and worker data subjects. Grounded by storyboards inspired by prior research literature proposing various wellbeing sensing technologies, we elicited stakeholders' envisioned harmful and beneficial impacts of wellbeing support in the workplace. We focused our study on understanding \textit{why} the workplace settings---with their complex organizational and social structures---surface unique concerns for workers, and \textit{how} the design of wellbeing technologies may interface with these challenges. Further, we engaged stakeholders in envisioning organizational, social, and technical mitigation strategies to address these concerns. 

\begin{acks}
This research was conducted at Microsoft Research. We thank our participants for their shared time and invaluable insights for the study, and the reviewers for their thoughtful and valuable feedback. We thank Michael Madaio, Samir Passi, Kenneth Holstein, Denae Ford Robinson, Vedant Das Swain, Matthew Jörke, Sachin Pendse, and researchers in the Microsoft Research FATE and HUE groups for providing feedback and additional resources to help conduct this work. 
\end{acks}

\bibliographystyle{ACM-Reference-Format}
\bibliography{references}

\end{document}